\documentclass[twocolumn,superscriptaddress,floatfix,preprintnumbers,prl ,nofootinbib,hyperref,longbibliography]{revtex4-1}
\usepackage{enumerate}
\usepackage{mathrsfs,amsmath,amssymb}
\usepackage{natbib}
\usepackage{graphicx}
\usepackage{cancel}
\usepackage[normalem]{ulem}
\usepackage[colorlinks, linkcolor=red, anchorcolor=blue, citecolor=blue, colorlinks=true, urlcolor=blue]{hyperref}
\hypersetup{colorlinks=true}
\usepackage{stackrel}
\usepackage{pgfplots}
\usepackage{subfigure}
\usepackage{makecell}
\usetikzlibrary{positioning}
\usetikzlibrary{snakes}

\newcommand{\prlsection}[2]{{\it\textbf{#1}{#2}}---}

\newcommand\be{\begin{equation}}
\newcommand\ee{\end{equation}}
\newcommand\bea{\begin{eqnarray}}
\newcommand\eea{\end{eqnarray}}

\begin{document}

\preprint{HEP-BNU-2022-0002}

\title{ Axion-like Dark Matter from the Type-II Seesaw Mechanism  }
\author{Wei Chao}
\email{chaowei@bnu.edu.cn}
\author{Mingjie Jin}
\email{jinmj507@gmail.com}
\author{Hai-Jun Li}
\email{lihaijun@bnu.edu.cn}
\author{Ying-Quan Peng}
\email{yqpenghep@mail.bnu.edu.cn}
\affiliation{Center for Advanced Quantum Studies, Department of Physics, Beijing Normal University, Beijing 100875, China}

\begin{abstract}
Although axion-like particles (ALPs) are popular dark matter candidates, their mass generation mechanisms as well as cosmic thermal evolutions are still unclear. In this letter, we propose a new mass generation mechanism of ALP during the electroweak phase transition in the presence of the type-II seesaw mechanism.  As ALP gets mass uniquely at the electroweak scale, there is a cutoff scale on the ALP oscillation temperature irrelevant to the specific mass of ALP, which is a distinctive feature of this scenario.  The ALP couples to the active neutrinos, leaving the matter effect of neutrino oscillations in a dense ALP environment as a smoking gun. As a by-product, the recent $W$-boson mass anomaly observed by the CDF collaboration is also quoted by the TeV-scale type-II seesaw.  We explain three kinds of  new physics phenomena  are  with one stroke. 

\end{abstract}

\maketitle

\prlsection{Introduction}{.} 
Various cosmological observations have confirmed the existence of cold dark matter (DM), which accounts for about 26.8\%~\cite{Planck:2018vyg} of the cosmic energy budget. 
However, the particle nature of DM still elude us.
Axion~\cite{Peccei:1977ur, Peccei:1977hh, Weinberg:1977ma, Wilczek:1977pj} is one of the most popular DM candidates  motivated by addressing the strong CP problem, with its mass induced by the QCD instanton and its relic abundance arising from the misalignment mechanism~\cite{Dine:1982ah, Preskill:1982cy, Abbott:1982af,Marsh:2015xka,DiLuzio:2020wdo,Co:2019jts}, which drives the coherent oscillation of axion field around the minimum of the effective potential.
Couplings of the axion to the standard model (SM) particles are model-dependent and there are three general types of QCD axion models, PQWW \cite{Peccei:1977ur, Peccei:1977hh}, KSVZ \cite{Kim:1979if, Shifman:1979if}, and DFSZ \cite{Dine:1981rt, Zhitnitsky:1980tq}, of which the PQWW axion is excluded by the beam-dump experiments \cite{Kim:1986ax,Faissner:1980qg,Faissner:1981hi} and other axion models can be detected via their couplings to photons or SM fermions.

To relax property constraints to the QCD axions, more general classes of axion-like particle (ALP) DM models \cite{Chikashige:1980ui,Gelmini:1980re,Wilczek:1982rv,Berezhiani:1990wn,Jaeckel:2013uva,Witten:1984dg,Conlon:2006tq,Cicoli:2012sz,Georgi:1981pu,Dias:2014osa,Choi:2009jt,Ringwald:2014vqa,Bauer:2018uxu,Reynoso:2022vrn} are proposed, with the mass ranging from $10^{-22}\,\rm eV$ to $\mathcal{O}(1)\, \rm GeV$ \cite{Marsh:2015xka,Bauer:2018uxu}, where the lower bound is  from the fuzzy DM~\cite{Hu:2000ke} and the upper bound is from the LHC limits.
The mass generation mechanism as well as the relic abundance of axion-like DM are blurred and indistinct since people usually pay more attention to the detection signal of ALP in various experiments via its coupling to photon \cite{Sikivie:1983ip, Raffelt:1987im, CAST:2007jps, Ehret:2010mh, HESS:2013udx, Fermi-LAT:2016nkz, CAST:2017uph, Gramolin:2020ict, Li:2020pcn, Salemi:2021gck, Reynes:2021bpe, Li:2022jgi}, $ a/f_a F\widetilde F$, where $a$ is the ALP field and $f_a$ is the ALP decay constant. 
It should be mentioned that the mass generation mechanism of the ALP is highly correlated with its interactions with the SM particles. So one cannot simply ignore these facts and directly apply the strategy of searching for QCD axion to detect the ALP.
This issue has been concerned recently and several novel approaches have been proposed to address the relic abundance of the light scalar DM,  such as the thermal misalignment mechanism \cite{Batell:2021ofv,Chun:2021uwr}, which supposes a feeble coupling between the DM and  thermal fermions. 
These attempts provide novel insights to the origin of ALP in the early Universe.

In this letter, we propose a new mechanism of generating the ALP mass during the electroweak phase transition  with the help of a Higgs triplet $\Delta$ with $Y=1$,  which is the seesaw particle in the type-II seesaw mechanism \cite{Lazarides:1980nt, Mohapatra:1980yp, Konetschny:1977bn, Cheng:1980qt, Magg:1980ut, Schechter:1980gr}.
Active neutrinos get Majorana mass as $\Delta$ develops a tiny but non-zero vacuum expectation value (VEV).  
We explicitly show that an ALP, which is the Goldstone boson arising from the spontaneous breaking of them global $U(1)_L$ symmetry, can get tiny mass through the quartic coupling with the Higgs triplet  and the SM Higgs doublet $\Phi$ whenever the global lepton number is explicitly broken by the term $\mu \Phi^T i\tau_2\Delta^\dag \Phi+\rm h.c.$
In such a scenario, symmetries break sequently: the $U(1)_L$ first breaks at  high energy scale resulting a massless ALP serving as dark energy, then electroweak symmetry is spontaneously broken leading the mass generation of the ALP, which begins to oscillate as its mass is comparable with the Hubble parameter.
We derive the relic density of ALP by investing its thermal evolution and solving its equation of motion (EOM) analytically.
To further investigate its signal,  we explicitly derive the interactions between ALP and SM particles,  which arise from the mixing of ALP with other CP-even particles.  We argue that neutrino oscillations in certain specific environment may be a smoking gun.  
As a by-product, we show that the recent $W$-boson mass anomaly observed by the CDF collaboration \cite{CDF:2022hxs,Lu:2022bgw,deBlas:2022hdk,Strumia:2022qkt,Fan:2022dck,Athron:2022qpo,
Bahl:2022xzi,Babu:2022pdn,Asadi:2022xiy,DiLuzio:2022xns,Athron:2022isz} can be addressed in the same model without conflicting with the LHC constraints.  



\prlsection{Framework}{.} We assume a complex scalar singlet $S$ carries  two units of lepton number charge and the $U(1)_L$ is spontaneously broken at high temperature when $S$ gets VEV. 
Besides, the type-II seesaw mechanism is required for the origin of neutrino mass and  $S$ couples to the Higgs triplet $\Delta$ and the SM Higgs doublet  $\Phi$ via the quartic interaction with a real coupling. 
The most general scalar potential is
\begin{eqnarray}
\begin{aligned}
V(S, \Phi,\Delta)
=&  V(\Phi, \Delta)-\mu^2_S (S^\dag S) +\lambda_6 (S^\dag S)^2  \\
&+  \lambda_7 (S^\dag S)(\Phi^\dag \Phi) + \lambda_8 (S^\dag S)\text{Tr}{(\Delta^\dag \Delta)} \\
&+ \mu \Phi^T i\tau_2\Delta^\dag \Phi+\lambda S \Phi^T i\tau_2\Delta^\dag \Phi+\rm{h.c.}\,, ~~~
\label{eq:pot}
\end{aligned}
\end{eqnarray}
where $V(\Phi, \Delta)$ is the most general potential for the type-II seesaw mechanism given in the Supplemental Material.
The quartic couplings $\lambda_{7, 8}$ are relevant for the thermal mass of $S$.  
It is obvious that $S$ may get non-zero VEV in the early Universe by assuming the small quartic couplings, which is consistent with experimental observations \cite{Schabinger:2005ei,Patt:2006fw,Pruna:2013bma,Lopez-Val:2014jva}, leaving the CP-odd component of $S$ as ALP.  
ALP is massless at the early time until the temperature drops down to the electroweak scale at which both $\Phi$ and $\Delta$ get non-zero VEVs. 
Then ALP acquires a tiny mass double suppressed by the VEV of the Higgs triplet and the tiny lepton-number-violating parameter $\mu$, which should be naturally small  accorded to the naturalness principle of t'Hooft \cite{tHooft:1979rat}.  

To analytically derive the mass of ALP, the $\Phi$, $\Delta$, and $S$ can be parametrized as
\begin{eqnarray}
\begin{aligned}
\Phi=\left[
\begin{array}{c}
\phi^+\\
\hspace{-1mm}\frac{v_\phi + \phi + i\chi}{\sqrt{2}} 
\end{array}\hspace{-1mm}\right], 
\Delta=
\left[
\begin{array}{cc}
\hspace{-1mm}\frac{\Delta^+}{\sqrt{2}}   & \Delta^{++}  \\
\hspace{-1mm}\Delta^0    & -\frac{\Delta^+}{\sqrt{2}}   
\end{array}\hspace{-1.5mm}\right],
S=\frac{v_s + \tilde{s} + i\tilde{a}}{\sqrt{2}}\label{eq:fields}, ~~~
\end{aligned}
\end{eqnarray}
where $\Delta^0=(v_\Delta+\delta + i \eta)/\sqrt{2}$ being the neutral component of the Higgs triplet,  the $v_\phi$, $v_\Delta$, and $v_s$ are the VEVs of $\Phi$, $\Delta$, and  $S$, respectively. 
After the electroweak symmetry breaking (EWSB), the remaining physical scalars are as follows, two charged scalar pairs $H^{\pm\pm}~\text{and}~H^{\pm}$, two CP-odd scalars $A~\text{and}~a$, and three CP-even scalars $h$, $H$, and $s$, whose masses may be obtained by unitary transformations to their squared mass matrices.   
The detailed procedures of diagonalization of all the scalar mass matrices are given in the Supplemental Material.  
Then the ALP mass in the CP-odd sector can be written as 
\begin{eqnarray}
m_{a}^2 \simeq { \sqrt{2}\mu v^2_\phi v_\Delta (v^2_\phi + 4 v^2_\Delta) \over 2 v^2_\phi (v^2_\Delta + v^2_s) + 8 v^2_\Delta v^2_s} \, .
\end{eqnarray}
In the limits $v_\Delta^2/v_\phi^2 \ll 1$ and $v_\Delta^2/v_s^2\ll1$,  one has $m^2_{a}\simeq \mu v^2_\phi v_\Delta/(\sqrt{2}v^2_s)$,
which is double suppressed by the parameters $v_\Delta$ and $\mu$ in the type-II seesaw mechanism.


\prlsection{ALP DM}{.}  
As discussed above, the ALP gets a tiny but non-zero mass via the type-II seesaw mechanism during the electroweak phase transition at the critical temperature $T_{\rm C} \simeq 160 \, \rm GeV$ \cite{DOnofrio:2014rug}.  
Neglecting the radiative corrections, the temperature-dependent ALP mass can be written as
\begin{align}
\label{ALP:mass}
m_{a}^2 (T)= 
\begin{cases}
\dfrac{\mu v^2_\phi (T) v_\Delta (T)}{\sqrt{2}f^2_{a} }\, ,\quad &T\leq T_{\rm C} \\
0 \, ,\quad   &T    >T_{\rm C}
\end{cases} 
\end{align}
where $f_a=v_s$, $v_\phi (T)$ and $v_\Delta (T)$ are the temperature-dependent VEVs of the SM Higgs and Higgs triplet, respectively. 
The EOM of the homogeneous ALP field $a$ (${a}\equiv\theta f_a$) in the FRW Universe can be written as
\cite{Dine:1982ah, Preskill:1982cy, Abbott:1982af}
\begin{align}
\label{ALPeom:dt}
\ddot{\theta}+3H(T)\dot{\theta}+ m_a^2(T)\theta = 0\, ,
\end{align}
where the dot denotes the derivative with the respect to time, and $H(T)\equiv\dot{R}/R$ is the Hubble parameter in terms of the scale factor $R$. 
In the radiation-dominated epoch, we have $H(T)=1/(2t)=1.66\sqrt{g_{*}(T)}T^2/m_{\rm pl}\,$,
where $g_{*}$ is the effective number of the degrees of freedom, and $m_{\rm pl}=1.221\times10^{19}\, \rm GeV$ being the Planck mass. 
The initial conditions are taken as $\theta(t_i) =\sqrt{\langle\theta_{a,i}^2\rangle}$ and $\dot{\theta}(t_i) = 0$, where the angle brackets denote the initial misalignment angle $\theta(t_i)$ averaged over $[ -\pi, \pi)$ \cite{DiLuzio:2020wdo}. 
The value of $\langle \theta_{a,i}^2\rangle$ depends on whether the $U(1)_L^{}$ breaking occurs before the inflation ends or after the inflation \cite{DiLuzio:2020wdo,Ringwald:2014vqa}.
 
\begin{figure}[t!]
\centering
\includegraphics[width=7.4cm]{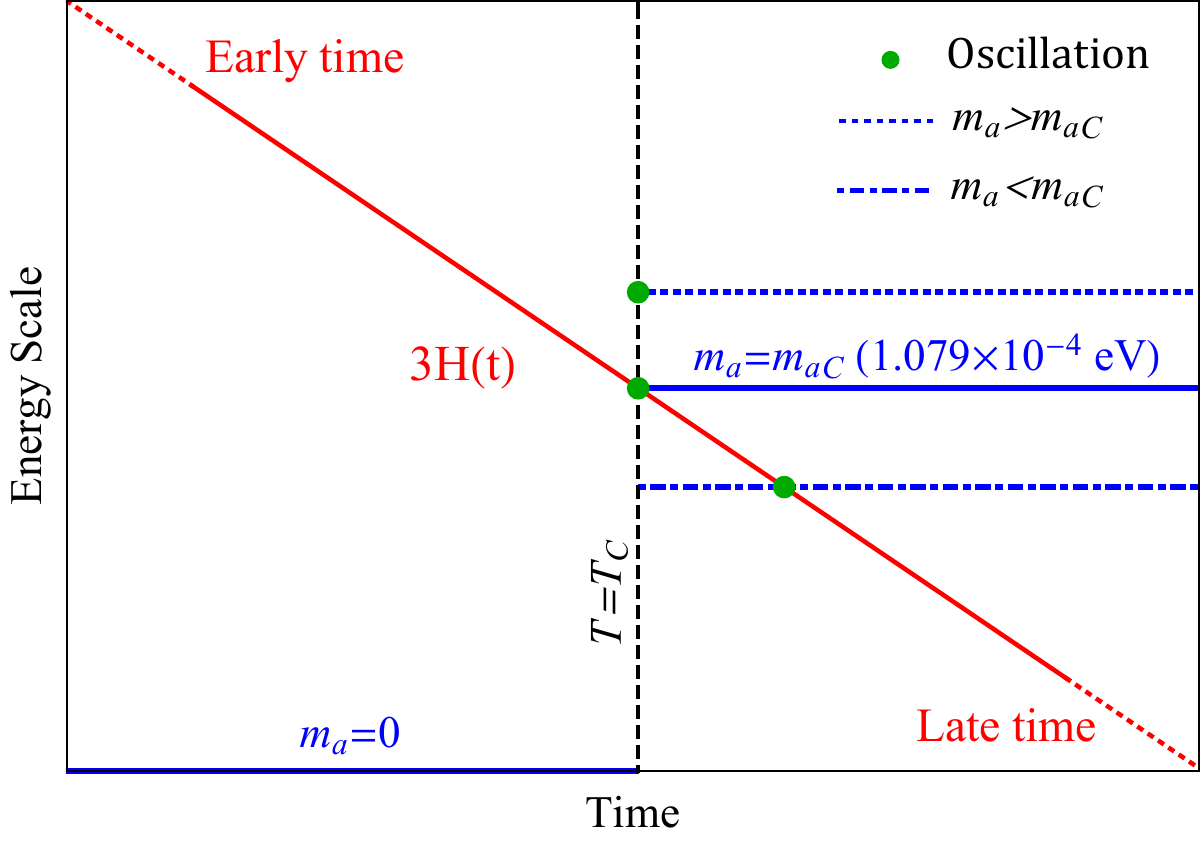}
\caption{The evolution of the energy scales for ALP mass $m_a$ (blue line) and the Hubble parameter (red line) as a function of the time. Three cases of $m_a$ are shown for comparisons. The green intersections represent the temperatures when the oscillation begins. The vertical dashed line represents the critical temperature ($T_{\rm C}$).}\label{Hubble:tem}
\end{figure}

In general, the ALP becomes dynamical and starts to oscillate when $m_a(T_{\rm{osc}})=3H(T_{\rm{osc}})$ \cite{Marsh:2015xka,DiLuzio:2020wdo,Co:2019jts}, where $T_{\mathrm{osc}}$ is the oscillation temperature. 
Before the EWSB, the ALP is massless and the angle $\theta$ remains a constant with the initial value $\theta(t)=\theta(t_i)$.  
Therefore, there is an upper bound on the oscillation temperature $T^{\rm max}_{\rm{osc}}\equiv T_{\rm C}$,  which leads to the existence of  a critical mass 
\begin{align}
\label{crit:mass}
{m_a}_{\rm C}=1.079\times10^{-4}\,{\rm eV}\, .
\end{align}
The oscillation temperature can be divided into two cases
\begin{align}
\label{crit:temp}
T_{\rm{osc}}
= \begin{cases}
T_*\, ,\quad &m_a< {m_a}_{\rm C}\\
T_{\rm C}\, ,\quad   &m_a \geq{m_a}_{\rm C}
\end{cases} 
\end{align}
where $T_*$ is derived from the condition $m_a=3H(T_*)$.
Eq.~(\ref{crit:temp}) implies that the traditional oscillation condition is only available to the case $m_a< {m_a}_{\rm C}$. 
For $m_a \geq{m_a}_{\rm C}$, the oscillation temperature is always equal to the critical temperature $T_{\rm C}$, as shown in Fig.~\ref{Hubble:tem}. 
Note that we use the parameter $3H$ instead of the Hubble parameter $H$ to better show the critical point given by  Eq.~(\ref{crit:temp}). 

\begin{figure}[t!]
\centering
\includegraphics[width=7.8cm]{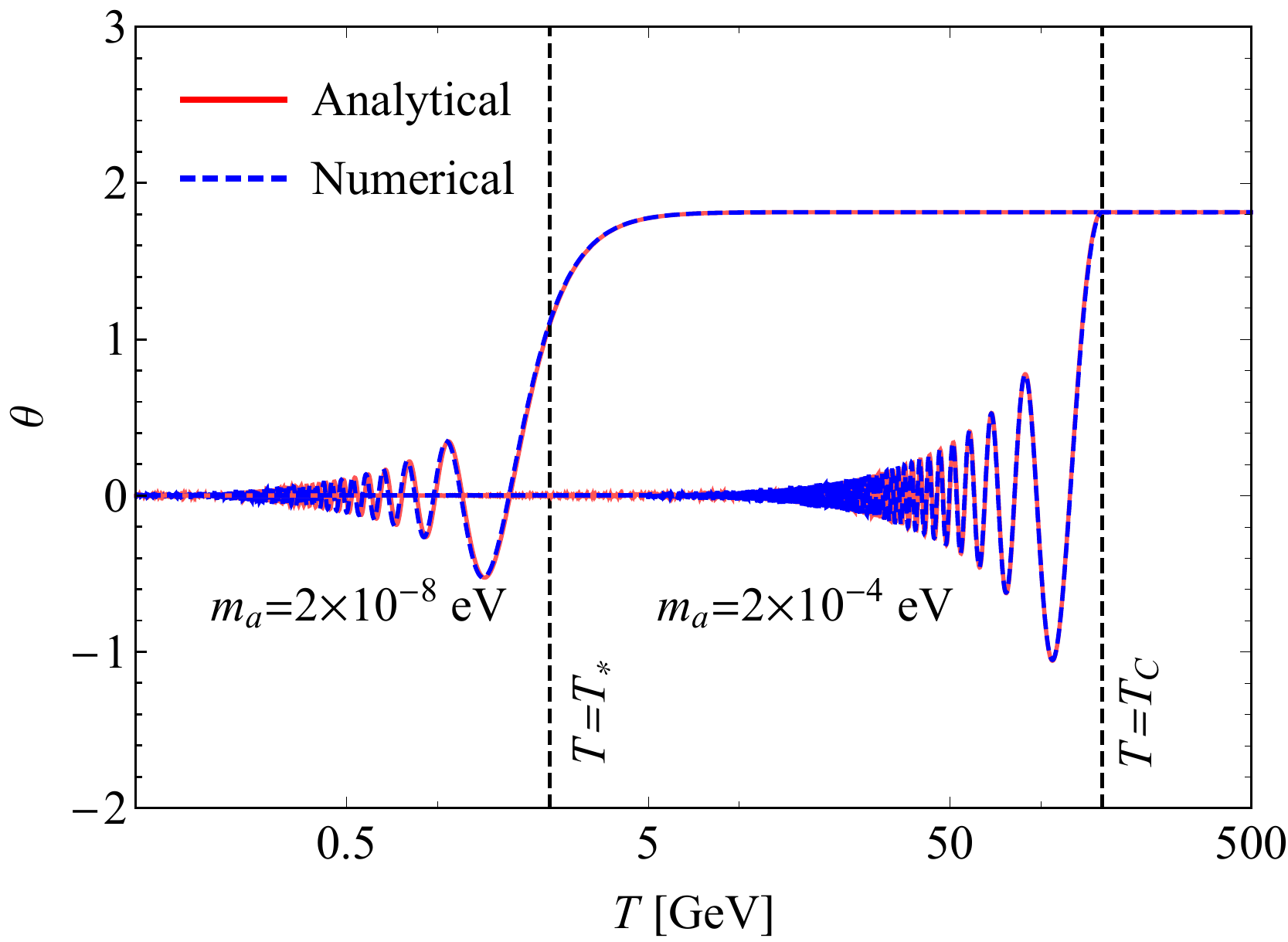}
\caption{The analytical (solid red) and numerical (dashed blue) evolution of $\theta$ as a function of $T$ for two benchmark ALP masses $m_a< {m_a}_{\rm C}$ ($m_a=2\times10^{-8}\, \rm eV$) and $m_a>{m_a}_{\rm C}$ ($m_a=2\times10^{-4}\, \rm eV$). The vertical dashed lines correspond to the oscillation temperatures.}
\label{theta:tem}
\end{figure}

We now investigate the evolution of the ALP, which is frozen at the initial value by the Hubble friction at early times (3$H>m_a$) and behaves as dark energy. 
As the temperature $T$ of the Universe drops to $T_{\rm{osc}}$ given by Eq.~(\ref{crit:temp}), the ALP  starts to oscillate with damped amplitude, and its energy density scales as $R^{-3}$, which is similar with the ordinary matter \cite{DiLuzio:2020wdo,Marsh:2015xka}, until the angle $\theta$ oscillates around the potential minimum of the ALP at the late time. 
The evolution of $\theta$ can be described by the {\it analytical} solution of EOM in the radiation dominated Universe when $H>H_E\sim10^{-28}\, {\rm eV}$ \cite{Baumann:2009ds,Marsh:2015xka}, where $H_E$ is the Hubble rate at the matter-radiation equality in $\Lambda{\rm CDM}$. 
The exact analytical expression is given in Sec.~B of the Supplemental Material.
Alternatively, we can also {\it numerically} solve Eq.~(\ref{ALPeom:dt}) with the given initial values. 
Here we consider the post-inflationary scenario and take the initial value as $\theta(t_i)=\pi/\sqrt{3}$ \cite{Ringwald:2014vqa,DiLuzio:2020wdo}. 
The analytical and numerical results are shown in Fig.~\ref{theta:tem} with the two benchmark ALP masses. 
We find that the numerical results of the evolution are  consistent with the analytical ones.

The energy density of ALP is $\rho_a(t) = \dot\theta^2(t)f_a^2/2 + m_a^2 (T) \theta^2(t) f_a^2/2$. 
Since the ratio of ALP number density to the entropy density is conserved, the ALP energy density at the present can be written as $\rho_a(T_0) \simeq \rho_a(R_{\rm osc}) \left(R_{\rm osc}/R\right)^3 = 1/2\, m_a(T_{\rm{osc}})m_a(T_0)f_a^2\left\langle\theta_{a,i}^2\right\rangle s(T_0)/s(T_{\rm osc})$ \cite{Marsh:2015xka,DiLuzio:2020wdo,Co:2019jts}, where $T_0$ is the CMB temperature at present, and $s=2\pi^2 g_{*s}T^3/45$ is the entropy density with  $g_{*s}$ the relativistic degrees of freedom of the entropy. 
The ALP mass is almost temperature-independent, which indicates $m_a(T_{\rm{osc}})=m_a(T_0)=m_a$, so the ALP energy density at present is  
\begin{align}
\label{rho:T0}
\rho_a(T_0) \simeq \frac{1}{2}m_a^2 f_a^2\left\langle \theta_{a,i}^2\right\rangle\frac{g_{*s}(T_0)}{g_{*s}(T_{\rm osc})}
\left(\frac{T_0}{T_{\rm osc}}\right)^3\, .
\end{align}
The relic density of ALP at present is defined as $\Omega_a h^2 = \left(\rho_a(T_0)/\rho_{c,0}\right)h^2$ \cite{Marsh:2015xka,DiLuzio:2020wdo}, 
where $\rho_{c,0}\equiv 3m_{\rm pl}^2H_0^2/\left(8\pi\right)$ is the critical energy density, $T_0=2.4\times10^{-4}\,\rm eV$, and $g_{*s}(T_0)=3.94$ \cite{Bauer:2017qwy}. 
Combining these parameters with Eq.~(\ref{rho:T0}), the relic density of ALP can be estimated as
\begin{eqnarray}
\begin{aligned}
\label{relica:T0}
\Omega_a h^2 = 
\begin{cases}
0.056\,\left\langle\theta_{a,i}^2\right\rangle\left(\dfrac{3.94}{g_{*s}(T_{*})}\right)\left( \dfrac{g_{*}(T_{*})}{3.36}\right)^{\frac{3}{4}}\\\times\left(\dfrac{f_a}{10^{13}\,{\rm GeV}}\right)^2\left(\dfrac{m_a}{10^{-7}\,{\rm eV}}\right)^{\frac{1}{2}}\,, ~m_a< {m_a}_{\rm C}\\
\\
0.0146\,\left\langle \theta_{a,i}^2\right\rangle\left(\dfrac{f_a}{10^{10}\,{\rm GeV}}\right)^2\left(\dfrac{m_a}{10^{-2}\,{\rm eV}}\right)^2\,,\\
m_a \geq{m_a}_{\rm C}
\end{cases} 
\end{aligned}
\end{eqnarray} 
Since the initial misalignment angle $\left\langle \theta_{a,i}^2\right\rangle^{1/2}\sim\mathcal{O} (1)$, the relic density is almost determined by the decay constant $f_a$ and its mass $m_a$. 
In Fig.~\ref{relicfa:ma}, we show the relic density $\Omega_a h^2$ as a function of $m_a$ with the four benchmark values of $f_a \sim \mathcal{O} (10^{10}-10^{13})\,\rm GeV$. 
The vertical black dotted line represents the critical mass $m_{a{\rm C}}$, on two sides of which the ALP density evolve differently. 
We find that there exists the allowed parameter space that may address the observed DM relic abundance, $\Omega_a h^2\simeq0.12$~\cite{Bauer:2017qwy,Planck:2018vyg}.

\begin{figure}[t]
\centering
\includegraphics[width=8.0cm]{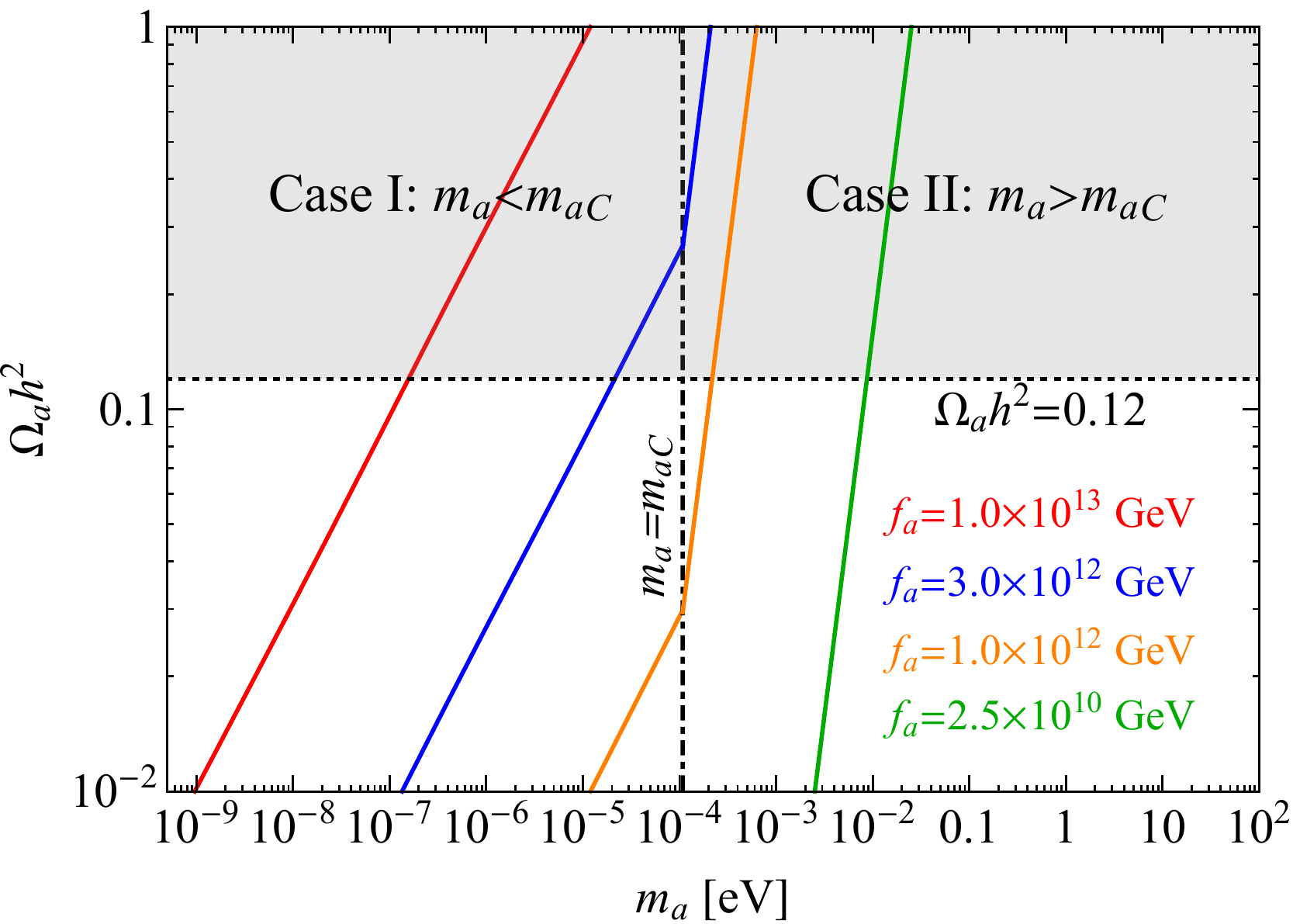}
\caption{The relic density $\Omega_a h^2$ as a function of $m_a$ for various $f_a$. The vertical dotted line represents the critical mass $(m_{a\rm C})$. The initial misalignment angle is taken as $\theta(t_i)=\pi/\sqrt{3}$. The gray region is excluded by the overabundance of DM.}
\label{relicfa:ma}
\end{figure}


\prlsection{ALP interactions}{.}  Now we investigate interactions of the ALP with ordinary matters including the Higgs and active neutrinos.  ALP  may couple to the SM Higgs as well as active neutrinos in forms $\lambda_{haa} h aa$ and $ \overline{ 
\nu_L^{C  }} i a\lambda_{a\overline{\nu}\nu}^{ } \nu_L^{}+ {\rm h.c.}$ with the couplings 
\begin{eqnarray}
\begin{aligned}
&\lambda_{haa}: -\lambda  U_{11} V_{13} V_{23} f_{a}+\frac{1}{2} \lambda  U_{21} V_{13}^2 f_{a}\,,\\
&\lambda_{a\overline{\nu}\nu}:  V_{23} m_\nu /v_\Delta\,,
\end{aligned}
\end{eqnarray} 
where  $U_{ij},V_{ij}$ ($i,j=1,2,3$) are the orthogonal matrices diagonalizing scalar matrices given in the Supplemental Material, and $m_\nu$ is the neutrino mass matrix in the flavor basis. 
The complete interactions of the ALP are listed in Table~IV of the Supplement Material. 
Given that the SM Higgs decays into two ALPs ($h \to aa$), the constraint of Higgs invisible decay from the LHC set an upper bound on the coupling $\lambda_{haa} < 1.536\, {\rm GeV}$ \cite{ParticleDataGroup:2020ssz}.
We have checked that the  coupling  predicted by this model always satisfy this constraint.

 \begin{figure}[t]
 \centering
 \includegraphics[width=7.5cm]{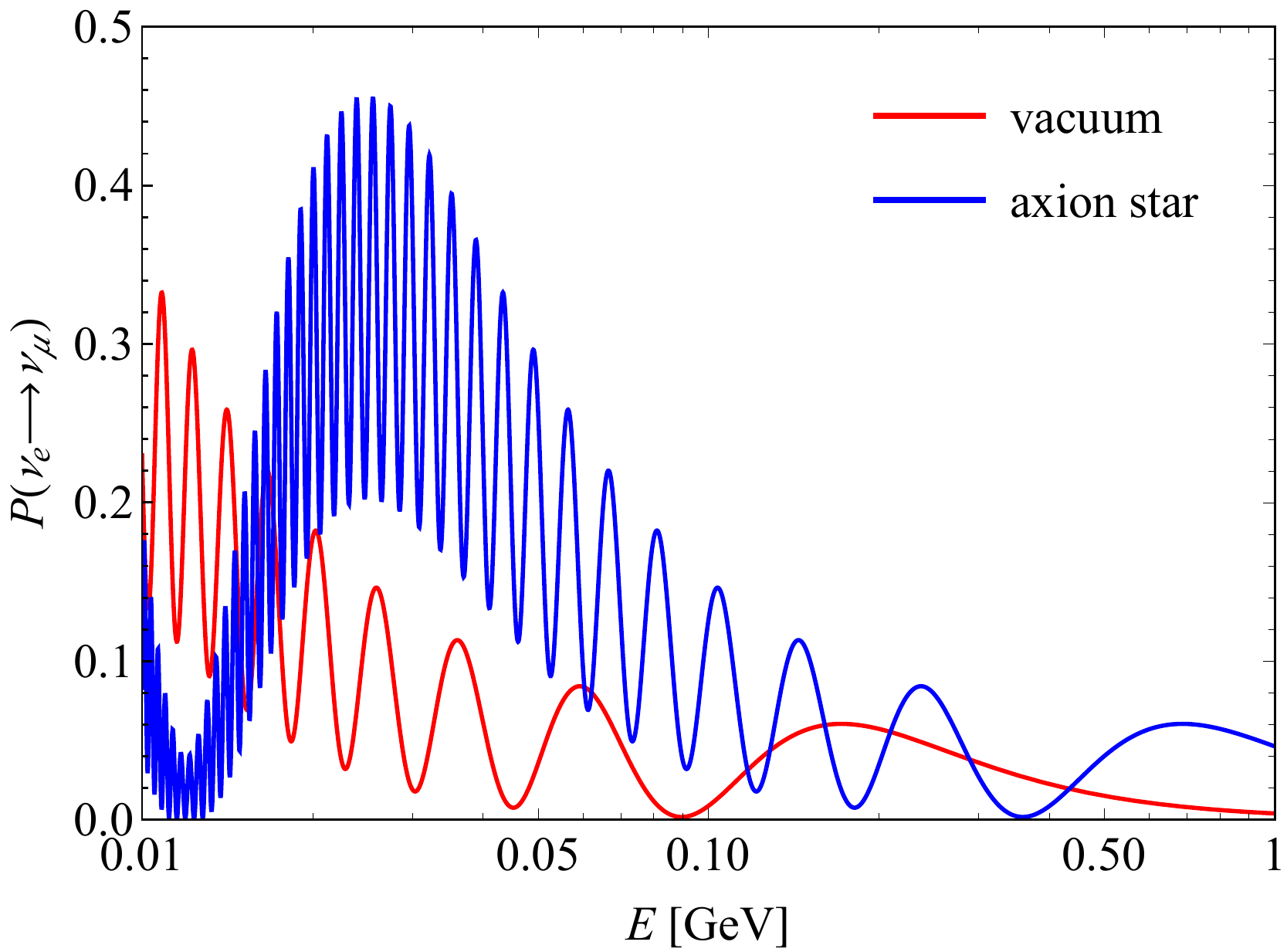}
 \caption{The transition probability $P$($\nu_e\to \nu_\mu$) as a function of the neutrino energy $E$.
 The red and blue lines stand for the neutrino oscillations in vacuum and dense axion stars, respectively. We take $m_a=10^{-5}\, \rm eV$, $f_a=10^{12}\, \rm GeV$, and $v_\Delta=1\, \rm MeV$ for axion, and take $M_a^{\rm dense}=13.6 M_{\odot}$ and $R_a^{\rm dense}=45.2\, \rm km$ for the dense axion star \cite{Chavanis:2017loo}. The three-flavor oscillation parameters are taken from Ref.~\cite{Esteban:2018azc}.}
 \label{neutrino:osc}
 \end{figure}


The interaction of ALP with active neutrinos may cause matter effect in neutrino oscillations. 
Since ALP is the classical field, the effective potential can be directly written as $V_{\rm eff}= i {\sqrt{2\rho_a }V_{23}m_a^{-1}} v_\Delta^{-1} \cos(m_a t ) \overline{\nu_L^C } m_\nu^{} \nu_L^{} + {\rm h.c.}$, which contributes an effective mass to active neutrinos  and  can be diagnonalized by the same unitary transformation as that in vacuum.   
In this case, the three-flavor neutrino oscillation amplitude can be written as 
\begin{eqnarray}
\begin{aligned}
A_{\alpha\to \beta } =&\sum_i \widehat{U}_{\beta i}\widehat{U}_{\alpha i}^*\exp\left[-i {m_i^2 x \over 2 E} \left( 1+ {\rho_aV_{23}^2 \over m_a^2 v_\Delta^2 }\right.\right.\\ 
&\left.\left.+ {\rho_aV_{23}^2 \cos 2 m_a x \over 2x m_a^3 v_\Delta^2} \right) \right]\,,
\label{neuoscillation}
\end{aligned}
\end{eqnarray} 
where $\widehat{U}_{\alpha i}$ is the matrix element of the PMNS matrix \cite{Maki:1962mu,Pontecorvo:1967fh}, $\alpha,\beta=\{e,\mu,\tau\}$, $i=\{1,2,3\}$, and $m_i$ is the mass of the $i$-th neutrino mass eigenstate. Notice that Eq.~(\ref{neuoscillation}) is same as the formula of neutrino oscillation in the vacuum up to the factor in the bracket. 

We find that it is difficult to probe this matter effect with a fixed $v_\Delta$ in vacuum, because of the low DM energy density $\rho_a$ and the super-small suppression factor $V_{23}$. 
The matter effect induced by this ALP-neutrino interaction becomes important only if the active neutrinos propagate in a dense celestial body, such as an axion star performed in~\cite{Braaten:2015eeu, Chavanis:2017loo}.
As an illustration, we show in Fig.~\ref{neutrino:osc} the neutrino oscillation probability  $P(\nu_e\to \nu_\mu)$  as a function of the neutrino energy in an axion star by setting  $\rho_a^{\rm dense}=6.97\times10^{19}\,{\rm g\,m^{-3}}$ \cite{Chavanis:2017loo}, which corresponds to the axion star of mass $M_a^{\rm dense}=13.6 M_{\odot}$ and radius $R_a^{\rm dense}=45.2\,\rm km$. In Fig.~\ref{neutrino:osc}, the matter effect induced by a dense axion star makes the neutrino oscillation spectrum different from that in the vacuum.

\prlsection{$W$ mass anomaly}{.}   Now we calculate the deviation of $W$-boson mass from the SM prediction at one-loop level within the framework of this model. 
In general, the expression of the $W$-boson mass $m_W$ can be parameterized as \cite{Hollik:1997pn,Awramik:2003rn}
\begin{align}
m_W^2 =
\frac{m_Z^2}{2}\left[1+\sqrt{1-\frac{4\pi\alpha_{\text{em}}}{\sqrt{2}G_Fm_Z^2}\left(1+\Delta r\right)}~\right]\, ,
\label{w-mass}
\end{align}
where $G_F$ is the Fermi constant, $\alpha_{\text{em}}$ is the fine-structure constant, and $\Delta r= \Delta\alpha_{\text{em}}-c_W^2/s_W^2\Delta\rho_{\rm loop}+\Delta r_{\text{rem}}$. 
The explicit expression of $\Delta r$ is given in the Supplemental Material.

\begin{figure}[t]
\centering
\includegraphics[width=8.0cm]{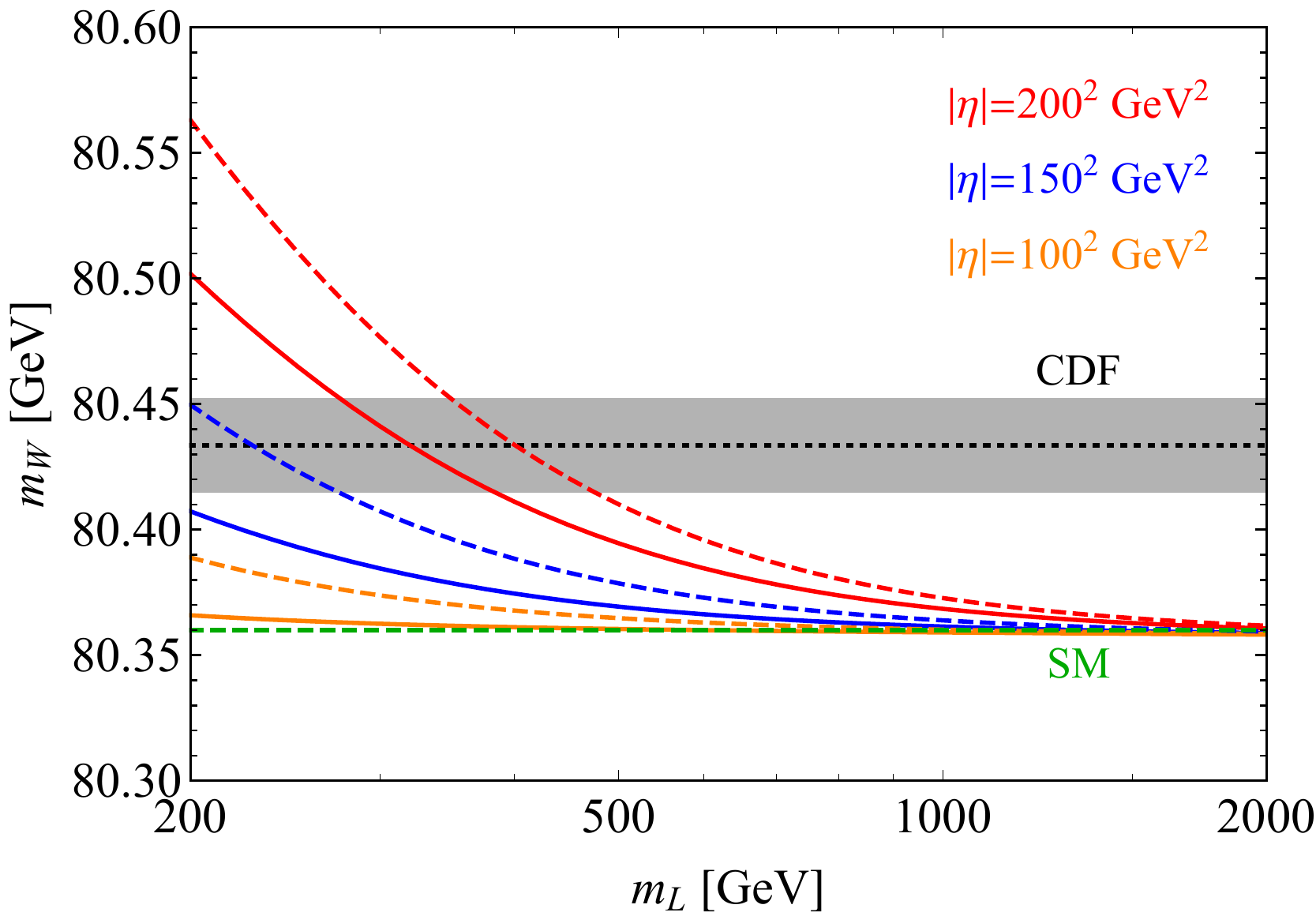}
\caption{The $m_W$ as a function of the lightest triplet-like Higgs $m_L$. Here we set $\mu=10^{-5}\,{\rm GeV}$, $v_\Delta=1\, \rm MeV$, $v_s=10^{13}\, \rm GeV$, and $m_{s}=1000\, \rm GeV$. Three typical values of $\eta$ are selected for comparisons. The dashed and solid lines correspond to the cases of $\eta<0$ and $\eta>0$, respectively. The dashed green line and the gray region represent the SM prediction \cite{Awramik:2003rn} and the recent 2$\sigma$ bound set by CDF \cite{CDF:2022hxs}, respectively.}
\label{deltarmw:mL}
\end{figure}

Both the Higgs triplet
~\cite{,Kanemura:2012rs,Aoki:2012jj,Kanemura:2022ahw,Cheng:2022jyi,Borah:2022obi,Heeck:2022fvl,Calibbi:2022wko,Popov:2022ldh,Chakrabarty:2022voz,Penedo:2022gej,Cheng:2022hbo,Berbig:2022hsm} and the scalar singlet
~\cite{Lopez-Val:2014jva,Sakurai:2022hwh} may contribute to $\Delta r$ and thus to the  $W$-boson mass. 
Given that $v_s$ is much larger than $v_\phi$ and  $v_\Delta$, it is reasonable to expect that the mixing angles $\alpha_2$ and $\alpha_3$ in CP-even sector are approximately zero ($\alpha_2,\alpha_3\simeq0$) as the scalar singlet is nearly decoupled from the other scalar fields.
The remaining $\alpha_1$ can be written as
\begin{align}
\alpha_1\simeq\arctan\Bigg[\frac{v_\phi v_\Delta (\lambda_4+\lambda_5)-{2M_\Delta^2 v_\Delta }/{v_\phi}}{{m_h}^2-M_\Delta^2}\Bigg]\ .
\label{eq:alpha123}
\end{align}
Here we take the coupling $\lambda_4=0$ \cite{Aoki:2012jj,Bahl:2022gqg} for simplicity.
The $M_\Delta^2$ and $\lambda_5$ correlated with the splitting of triplet mass spectrum are obtained from the limits of $v_\Delta^2/v_\phi^2, \, v_\Delta^2/v_s^2\ll1$ as
\begin{align}
\lambda_5\simeq\frac{4m_A^2 - 4 m_{H^{+}}^2 }{v^2_\phi}\simeq\frac{4m_{H^{+}}^2-4m_{H^{++}}^2}{v^2_\phi}\,, 
\end{align}
and $M_\Delta^2\simeq m_A^2$. 

We denote the lightest mass of $m_{H^{++}}$, $m_{H^+}$, and $m_{A}$ as the variable $m_L$ and show in Fig.~\ref{deltarmw:mL} the improved $m_W$ as the function of $m_L$ for various $\eta$, which is defined as $\eta \equiv m_{H^{++}}^2-m_{H^+}^2=m_{H^{+}}^2-m_{A}^2$. 
Three typical values of  $|\eta|=(100)^2\,{\rm GeV}^2$, $(150)^2\,{\rm GeV}^2$, and $(200)^2\,{\rm GeV}^2$ are selected for comparisons.
The dashed and solid lines correspond to the cases of $\eta<0$ and $\eta>0$, respectively.
Notice that  $m_W$ would asymptotically approach to the SM prediction with the increase of $m_L$ due to the decoupling of Higgs triplet. 
However, when $m_L$ is lower than $2000\, \rm GeV$, for different $\eta$, $m_W$ increases diversely with the decrease of $m_L$, and some curves can reach the range of the CDF measurement~\cite{CDF:2022hxs}. In this case, the CDF anomaly can be feasibly explained by taking $|\eta|\in[(150)^2,(200)^2]\,{\rm GeV}^2$ for $m_L \lesssim 500\, \rm GeV$.


\prlsection{Summary}{.} 
%
In this letter, we have proposed a new mass generation mechanism of ALP from the type-II seesaw mechanism that give rise to the active neutrino Majorana masses. 
The typical oscillation temperature of ALP shows a cut-off at the critical temperature of the EWSB, which is the typical trappings of this kind of ALP.  
Although the ALP does not couple to the diphoton, it might be detected in future neutrino oscillation experiments due to the matter effect induced by the ALP-neutrino interactions. 
Finally, we show that the $W$-mass anomaly observed by the CDF  collaboration can be explained by the TeV-scale type-II seesaw.  All these observations make three  different kinds of new physics phenomena tightly connected with each other in a single model.


\prlsection{Acknowledgments}{.} 
The authors would like to thank Sara Khatibi for helpful discussions.
This work was supported by the National Natural Science Foundation (NNSF) of China (Grants No.~11775025 and No.~12175027).

\bibliography{references}

\clearpage
\newpage
\maketitle
\onecolumngrid
\begin{center}
\textbf{\large Axion-like Dark Matter from the Type-II Seesaw Mechanism} \\
\vspace{0.1in}
{ \it \large Supplemental Material}\\
\vspace{0.05in}
{Wei Chao, Mingjie Jin, Hai-Jun Li, and Ying-Quan Peng}
\end{center}
\onecolumngrid
\setcounter{figure}{0}
\setcounter{table}{0}
\setcounter{section}{0}
\setcounter{page}{1}
\makeatletter
\renewcommand{\theequation}{S\arabic{equation}}
\renewcommand{\thefigure}{S\arabic{figure}}

The Supplemental Material is organized as follows. 
In Sec.\ref{Singlet_Triplet}, we present the model in detail.
The analytical solution of the EOM for ALP is given in Sec.\ref{analytic:EOM}. In Sec.\ref{decay:Higgs_Maj}, we calculate the branching fraction of the decay process $h\to a a$. In Sec.\ref{W_mass}, we discuss the calculation of $W$-boson mass anomaly. 
Finally, the gauge-scalar interactions, the ALP interactions, and the gauge boson self-energies are listed in Secs.\ref{int:gauge_scalar},\ref{int:Maj_scaneu}, and \ref{self_energies}, respectively.

\subsection{The Singlet-Triplet Model}
\label{Singlet_Triplet}

The relevant Lagrangian is given by
\begin{align}
\mathcal{L}_{\text{SHT}}=(\partial_\mu S)^\dagger (\partial^\mu S) + (D_\mu \Phi)^\dagger (D^\mu \Phi)+\text{Tr}[(D_\mu \Delta)^\dagger (D^\mu \Delta)]- V(S,\Phi,\Delta)+ \mathcal{L}_{\rm Yukawa}\,,
\end{align}
where the covariant derivatives are defined as
\begin{align}
D_\mu \Phi=\left(\partial_\mu+i\frac{g}{2}\tau^aW_\mu^a+i\frac{g'}{2}B_\mu\right)\Phi\,, \quad
D_\mu \Delta=\partial_\mu \Delta+i\frac{g}{2}\bigg[\tau^aW_\mu^a,\Delta\bigg]+ig'B_\mu\Delta\,.
\end{align}
The general form of the scalar potential $V(S, \Phi,\Delta)$ is given by
\begin{align}
V(S, \Phi,\Delta) 
&= {-\mu^2_\Phi (\Phi^\dag \Phi) +\mu_\Delta^2 \text{Tr}(\Delta^\dag\Delta) -\mu^2_S (S^\dag S)} + \lambda_1 (\Phi^\dag \Phi)^2 \notag \\
&~~~~~+\lambda_2 \left[\text{Tr}(\Delta^\dag\Delta)\right]^2+\lambda_3\text{Tr}[(\Delta^\dag\Delta)^2] +\lambda_4(\Phi^\dag\Phi)\text{Tr}(\Delta^\dag\Delta) +\lambda_5\Phi^\dag\Delta\Delta^\dag\Phi \notag\\
&~~~~~+ \lambda_6 (S^\dag S)^2+  \lambda_7 (S^\dag S)(\Phi^\dag \Phi) + \lambda_8 (S^\dag S)\text{Tr}{(\Delta^\dag \Delta)} \notag \\
&~~~~~+ \mu \Phi^T i\tau_2\Delta^\dag \Phi+\lambda S \Phi^T i\tau_2\Delta^\dag \Phi+\text{h.c.}\,,
\label{eq:pot}
\end{align}
and $\mathcal{L}_{\rm Yukawa}$ is the Yukawa interaction of  left-handed lepton doublets \cite{Lazarides:1980nt, Mohapatra:1980yp, Konetschny:1977bn, Cheng:1980qt, Magg:1980ut, Schechter:1980gr},
\begin{align}
-\mathcal{L}_{\rm Yukawa}&=y_{\alpha\beta}\overline{\ell_L^{\alpha c}}i\tau_2\Delta \ell_L^\beta+\text{h.c.}\,, 
\label{nu_yukawa}
\end{align}
where $y_{\alpha\beta}$ denotes the $3\times 3$ complex symmetric matrix, and $\ell_L^\alpha=( \nu_L^\alpha,e_L^\alpha) ^{T}$ is the left-handed lepton doublet with $\alpha=\{e,\mu,\tau\}$. From Eq.~(\ref{nu_yukawa}) we know that $\Delta $ carries a lepton number charge of $-2$. The scalar fields $\Phi$, $\Delta$, and $S$ can be parametrized as
\begin{align}
\Phi=\left(
\begin{array}{c}
\phi^+\\
\frac{v_\phi + \phi + i\chi}{\sqrt{2}} 
\end{array}\right)\,,
\quad
\Delta=
\left(
\begin{array}{cc}
\frac{\Delta^+}{\sqrt{2}}   & \Delta^{++} \\
\Delta^0    & -\frac{\Delta^+}{\sqrt{2}}   
\end{array}
\right)\,,
\quad
S=\frac{v_s + \tilde{s} + i\tilde{a}}{\sqrt{2}}\,,
\end{align}
where $\Delta^0=(v_\Delta +\delta + i \eta)/\sqrt{2}$, and $v^2\equiv v_\phi^2+2v_\Delta^2\simeq$ $(246\, {\rm GeV})^2$. 
The $v_\phi$, $v_\Delta$, and $v_s$ are the VEVs of the Higgs doublet, the Higgs triplet, and the scalar singlet, respectively.  After the Higgs triplet acquires a VEV $v_\Delta$, Eq.~(\ref{nu_yukawa}) gives rise to the mass matrix of active neutrinos,
\begin{align}
(m_\nu)_{\alpha\beta}= y_{\alpha\beta}v_\Delta /\sqrt{2}\,.  
\label{eq:mn}
\end{align}
At the tree level, the $W$ boson and the $Z$ boson obtain masses through Higgs mechanism,
\begin{align}
m_W^2 = \frac{g^2}{4}\left(v_\phi^2+2v_\Delta^2\right)\,,\quad 
m_Z^2 =\frac{g^2}{4\cos^2\theta_W}\left(v_\phi^2+4v_\Delta^2\right)\,. 
\label{mV}
\end{align}
The electroweak $\rho$ parameter can slightly deviate from 1, $\rm i.e.$ \cite{Peskin:1991sw},
\begin{align}
\rho \equiv \frac{m_W^2}{m_Z^2\cos^2\theta_W}=\frac{1+\frac{2v_\Delta^2}{v_\phi^2}}{1+\frac{4v_\Delta^2}{v_\phi^2}}\,.
\label{rho_triplet}
\end{align}
Actually, the experimental measurement of the $\rho$ parameter gives $\rho^{\text{exp}}=1.0002\pm{0.0009}$ \cite{ParticleDataGroup:2020ssz}, which implies that $v_\Delta\lesssim7\, \rm GeV$ \cite{Kanemura:2022ahw} according to Eq.~(\ref{rho_triplet}). The physical scalar sectors are obtained by rotating the weak eigenstates of the scalar fields with the following orthogonal transformations
\begin{align}
\left(
\begin{array}{c}
G^\pm\\
H^\pm
\end{array}\right)=
\mathcal{R}(\beta)
\left(
\begin{array}{c}
\phi^\pm\\
\Delta^\pm
\end{array}\right)\,,\quad
\left(
\begin{array}{c}
G\\
A\\
a
\end{array}\right)=
\mathcal{V}({\beta'_1,\beta'_2,\beta'_3})
\left(
\begin{array}{c}
\chi\\
\eta\\
\tilde{a}
\end{array}\right)\,,\quad
\left(
\begin{array}{c}
h\\
H\\
s
\end{array}\right)=
\mathcal{U}({\rm \alpha_1,\alpha_2,\alpha_3})
\left(
\begin{array}{c}
\phi\\
\delta\\
\tilde{s}
\end{array}\right)\,, 
\label{mix:angle}
\end{align}
where the expressions of the orthogonal matrices $\mathcal{R}(\beta)$, $\mathcal{V}({\beta'_1,\beta'_2,\beta'_3})$, and $\mathcal{U}({\rm \alpha_1,\alpha_2,\alpha_3})$ can be found in Ref.~\cite{Arhrib:2018qmw}. The mixing angles  $\beta$ and $\beta_i$  are
\begin{align}
\tan\beta=\frac{\sqrt{2}v_\Delta}{v_\phi}\,,\quad \tan\beta'_1 = \frac{2v_\Delta}{v_\phi}\,, \quad
\tan\beta'_2=0\,, \quad \tan2\beta'_3= \frac{-2\lambda v_\Delta v_s v_\phi \sqrt{v^2_\phi + 4 v^2_\Delta}}{v^2_\phi \left(- \lambda v^2_\Delta + \lambda v^2_s + \sqrt{2} \mu v_s\right) + 4 v^2_\Delta v_s \left(\sqrt{2}\mu + \lambda v_s\right)}\,.
\end{align}
The masses of charged and CP-odd physical states are 
\begin{align}
&m_{H^{++}}^2=M_\Delta^2-\lambda_3v_\Delta^2-\frac{\lambda_5}{2}v_\phi^2\simeq M_\Delta^2-\frac{\lambda_5}{2}v_\phi^2 \,,\quad
m_{H^+}^2= \left(M_\Delta^2-\frac{\lambda_5}{4}v_\phi^2\right)\left(1+\frac{2v_\Delta^2}{v_\phi^2}\right)\simeq M_\Delta^2-\frac{\lambda_5}{4}v_\phi^2\,, \notag\\
&m_A^2 =M_\Delta^2\left(1+\frac{4v_\Delta^2}{v_\phi^2} + \frac{v^2_\Delta}{v^2_s} \right)\simeq M_\Delta^2,\quad m_a^2=\frac{\sqrt{2}\mu v^2_\phi v_\Delta (v^2_\phi + 4 v^2_\Delta)}{2 v^2_\phi (v^2_\Delta + v^2_s) + 8 v^2_\Delta v^2_s}\simeq \frac{\mu v^2_\phi v_\Delta }{\sqrt{2}v_s^2}\,, \label{mde}
\end{align}
where $M_\Delta^2=\frac{\mu v^2_\phi}{\sqrt{2}v_\Delta} + \frac{\lambda v_s v^2_\phi}{2 v_\Delta}\approx \frac{\lambda v_s v^2_\phi}{2 v_\Delta}$. Here we defines $\eta \equiv m_{H^{++}}^2-m_{H^+}^2=m_{H^{+}}^2-m_{A}^2$.

Given that the value of $v_s$ is much larger than that of $v_\phi$ and  $v_\Delta$, it is reasonable to expect that the mixing angles $\alpha_2$ and $\alpha_3$ in CP-even sector are approximately zero ($\alpha_2,\alpha_3\simeq0$) because the scalar singlet is nearly decoupled from the other scalar fields. The remaining $\alpha_1$ can be written as 
\begin{align}
\alpha_1 \simeq\arctan\left[\frac{v_\phi v_\Delta \left(\lambda_4+\lambda_5\right)-{2M_\Delta^2 v_\Delta }/{v_\phi}}{{m_h}^2-M_\Delta^2}\right]\, . \quad
\label{eq:alpha123}
\end{align}
Here  we take  $\lambda_4=0$~\cite{Aoki:2012jj,Bahl:2022gqg}, and  $\lambda_5$  which is correlated with the splitting of triplet mass spectrum  is given as
\begin{align}
\lambda_5\simeq\frac{4m_A^2 - 4 m_{H^{+}}^2 }{v^2_\phi}\simeq\frac{4m_{H^{+}}^2-4m_{H^{++}}^2}{v^2_\phi}. 
\end{align}
In this case ($\alpha_2,\alpha_3\simeq0$), the mass eigenvalues of CP-even physical states are
\begin{align}
m_{h}^2&=v_\phi \left(2\lambda_1 v_\phi - {\rm tan}\alpha_1     \left(2 M_\Delta^2 v_\Delta/v_\phi^2 - v_\Delta (\lambda_4 + \lambda_5) \right)\right) \,,\notag\\
m_{H}^2&= v_\phi \left(2\lambda_1 v_\phi + {\rm cot}\alpha_1 \left(2 M_\Delta^2 v_\Delta/v_\phi^2 - v_\Delta (\lambda_4 + \lambda_5) \right)\right)\,, \notag\\
m_s^2 &= \frac{\lambda v_\Delta v_\phi^2}{2 v_s} + 2 \lambda_6 v_s^2 \,. \label{mde}
\end{align}

\subsection{Analytical solution of the EOM}
\label{analytic:EOM}

To solve the EOM for ALP given by Eq.~(5) analytically, we take the initial conditions as \cite{Ringwald:2014vqa,DiLuzio:2020wdo}
\begin{align}
\theta(t_i)=\frac{\pi}{\sqrt{3}}\,,\quad \dot{\theta}(t_i)=0\,, 
\end{align}
then the analytical solution is 
\begin{align}
	\theta(t)=& -\pi\bigg[-2m_at_iJ_{\frac{1}{4}}(m_at)Y_{-\frac{3}{4}}(m_at_i)
	+2m_at_iY_{\frac{1}{4}}(m_at)J_{-\frac{3}{4}}(m_at_i)
	-Y_{\frac{1}{4}}(m_at)J_{\frac{1}{4}}(m_at_i) -2m_at_iY_{\frac{1}{4}}(m_at)_{\frac{5}{4}}(m_at_i)\nonumber\\
	&+J_{\frac{1}{4}}(m_at)Y_{\frac{1}{4}}(m_at_i)
	+2m_at_iJ_{\frac{1}{4}}(m_at)Y_{\frac{5}{4}}(m_at_i) \bigg] \bigg/
	\bigg\{2\sqrt{3}t^{\frac{1}{4}}t_i^{\frac{3}{4}}\bigg[J_{\frac{1}{4}}(m_at_i)Y_{-\frac{3}{4}}(m_at_i) 
	-J_{-\frac{3}{4}}(m_at_i)Y_{\frac{1}{4}}(m_at_i)\nonumber\\&
	+J_{\frac{5}{4}}(m_at_i)Y_{\frac{1}{4}}(m_at_i)
	-J_{\frac{1}{4}}(m_at_i)Y_{\frac{5}{4}}(m_at_i)\bigg] \bigg\}\,,
\end{align}
where $J_n$ and $Y_n$ are the Bessel functions of rank-n. 
By taking $t_i\to0$ \cite{Chun:2021uwr} and $t_i=t_{\rm C}$ (the critical time when ALP starts to oscillate at $T_{\rm C}$), the analytical evolution  is shown  in Fig.~2  as the red curves, as a comparison of the numerical simulations (blue curves).

\subsection{The decay of SM Higgs into ALPs}
\label{decay:Higgs_Maj}

\begin{figure}[h]
\centering
\includegraphics[width=8.5cm]{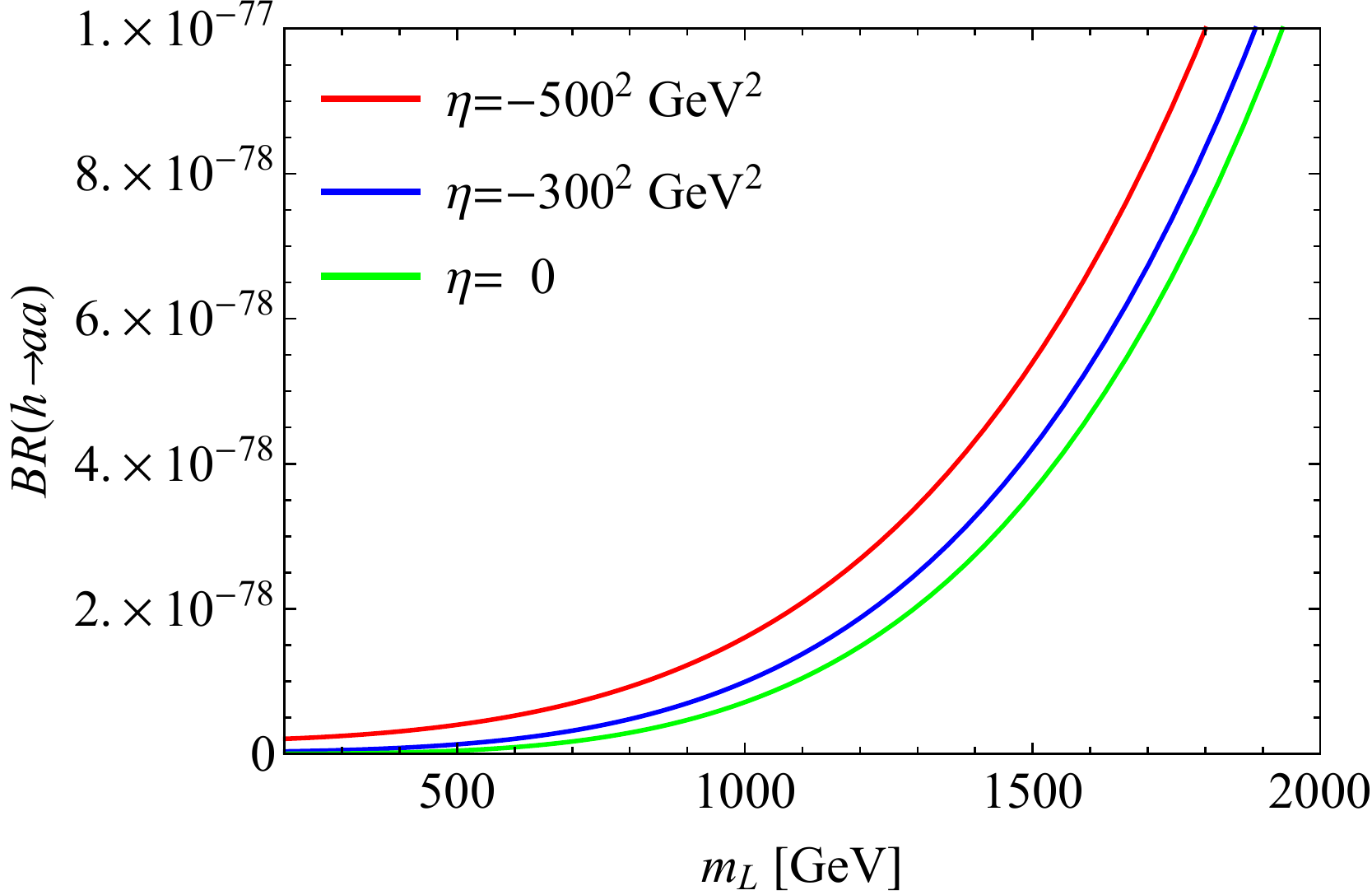}
\caption{The branching fraction of $h\to aa$ as a function of $m_L$.}
\label{branch:haa}
\end{figure}

The SM Higgs can decay into ALPs where the relevant coupling is listed in Table~\ref{fr4}, and the decay width is estimated to be
\begin{align}
\label{width:haa}
\Gamma_{h\to aa}=\frac{\lambda^2_{haa}}{8\pi m_h}\left(1-\frac{4m^2_a}{m_h^2}\right)^{{1}/{2}}\, ,
\end{align}
where $\lambda_{haa}\equiv-\lambda  U_{11} V_{13} V_{23} v_{s}+\frac{1}{2} \lambda  U_{21} V_{13}^2 v_{s}$ with $\lambda \approx\frac{2 v_\Delta M_\Delta^2}{v_s v^2_\phi}$, and we neglect the $v_\Delta$ and $v_\phi$ terms. The total width of the SM Higgs is $\Gamma_{h}=3.2\, \rm MeV$ \cite{ParticleDataGroup:2020ssz}. As discussed in Fig.~5, we take $\mu=10^{-5}\,{\rm GeV}$, $v_\Delta=1\, \rm MeV$, $v_s=10^{13}\, \rm GeV$, and $m_{s}=1000\, \rm GeV$. Then the branching fraction of the decay $h\to aa$ as a function of $m_L$ is shown in Fig.~\ref{branch:haa}. Note that the branching fraction is of order $\sim\mathcal{O}(10^{-78})$, which implies that this decay process is almost impossible to be detected by current experiments.

\subsection{One-loop radiative corrections to $W$-boson mass}
\label{W_mass}

In this section, we calculate the $W$-boson mass at the one-loop level. First of all, the expression of $m_W$ is given by \cite{Hollik:1997pn,Awramik:2003rn}
\begin{align}
m_W^2 =\frac{m_Z^2}{2}\left[1+\sqrt{1-\frac{4\pi\alpha_{\text{em}}}{\sqrt{2}G_Fm_Z^2}\left(1+\Delta r\right)}~\right]\, ,
\label{w-mass}
\end{align}
where $\Delta r$ is defined as
\begin{align}
\Delta r &= \Delta\alpha_{\text{em}}-\frac{c_W^2}{s_W^2}\Delta\rho_{\rm loop}+\Delta r_{\text{rem}}\, , 
	\label{delr3}
\end{align}
with~\cite{Kanemura:2012rs,Aoki:2012jj,Kanemura:2022ahw}
\begin{align}
\Delta \alpha_{\text{em}} = \Pi_{\gamma\gamma}'(0)-\Pi_{\gamma\gamma}^\prime(m_Z^2)\, ,
\label{del_alpha}
\end{align}
\begin{eqnarray}
\begin{aligned}
\Delta \rho_{\rm loop}  =\frac{\Pi_{ZZ}(0)}{m_Z^2}-\frac{\Pi_{WW}(0)}{m_W^2} + \frac{2s_W}{c_W}\frac{\Pi_{Z\gamma }(0)}{m_Z^2}\, ,
\label{del_rho}
\end{aligned} 
\end{eqnarray}
\begin{eqnarray}
\begin{aligned}
\Delta r_{\text{rem}}=\frac{c_W^2}{s_W^2}\left[\frac{\Pi_{ZZ}(0)}{m_Z^2}-\frac{\text{Re}\left[\Pi_{ZZ}(m_Z^2)\right]}{m_Z^2}\right]
+\left(1-\frac{c_W^2}{s_W^2}\right)\left[\frac{\Pi_{WW}(0)}{m_W^2}-\frac{\text{Re}\left[\Pi_{WW}(m_W^2)\right]}{m_W^2}\right]
+\Pi_{\gamma\gamma}'(m_Z^2)+\delta_{\rm VB}\, . \quad
\label{del_rem}
\end{aligned} 
\end{eqnarray}
The explicit expressions for the gauge boson self-energies $\Pi_{WW}$, $\Pi_{\gamma\gamma}$, and $\Pi_{Z\gamma }$ are listed in Sec.~\ref{int:gauge_scalar}. 
The $\delta_{\rm VB}$ denotes the contribution from the vertex and box radiative corrections, which are calculated in Refs.~\cite{Blank:1997qa,Hagiwara:1994pw}. 
Other input experimental values related to electroweak parameters are \cite{CDF:2022hxs,ParticleDataGroup:2020ssz}
\begin{align}
&~\alpha_{\text{em}}^{-1} = 137.035999\, ,\quad
G_F = 1.1663787\times 10^{-5} \,\text{GeV}^{-2}\,,\quad\notag\\
&~\hat{s}_W^2=0.23121\, ,\quad\quad~~
m_h=125.10\,\text{GeV}\,,\quad\notag\\
&~m_Z =91.1876 \,\text{GeV}\, ,\quad
m_W^{\text{CDF}} = 80.4335\pm0.0094\,\text{GeV}\, .
\end{align}

\begin{figure}[h]
\centering
\includegraphics[width=8.5cm]{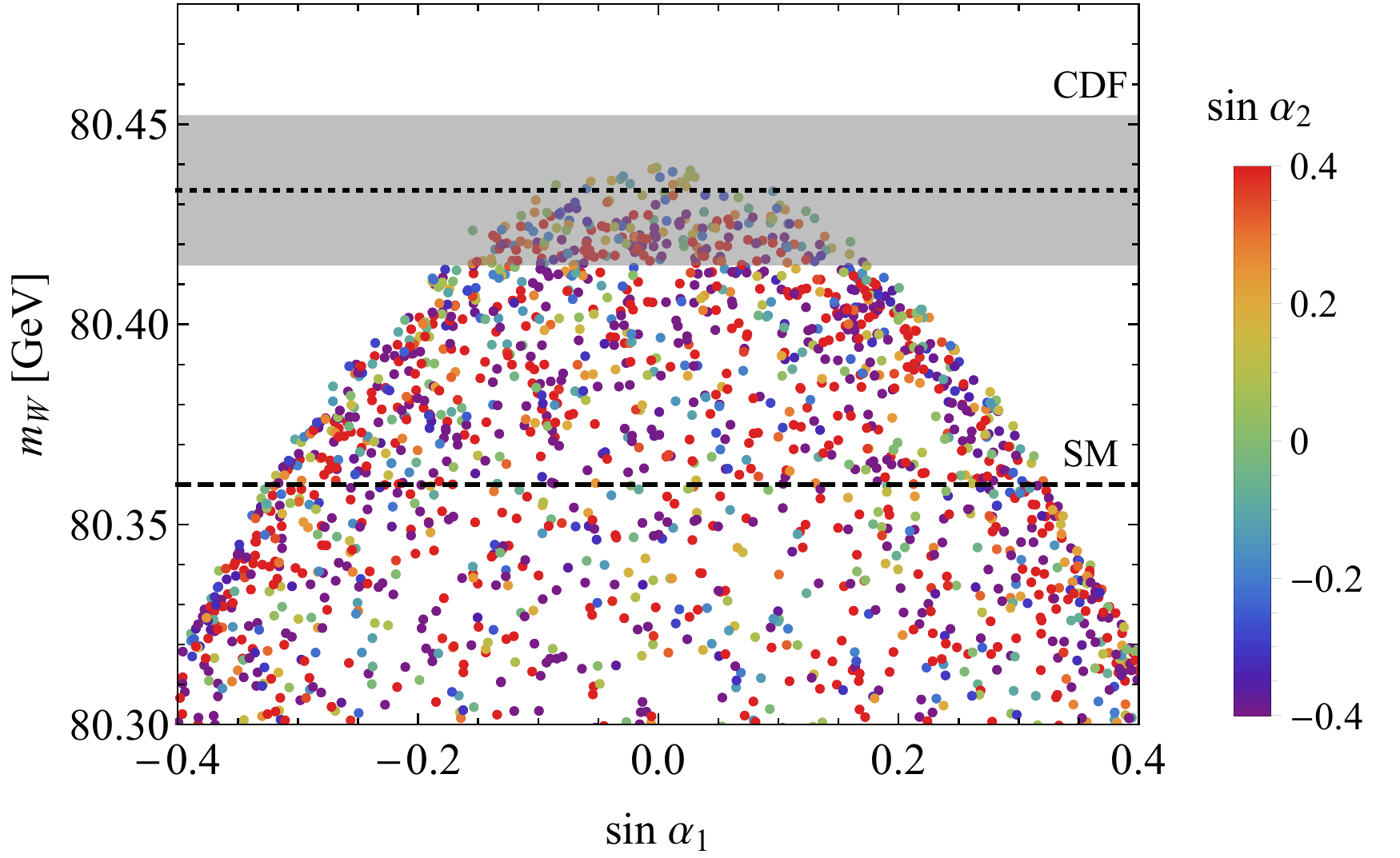}
\quad
\includegraphics[width=8.5cm]{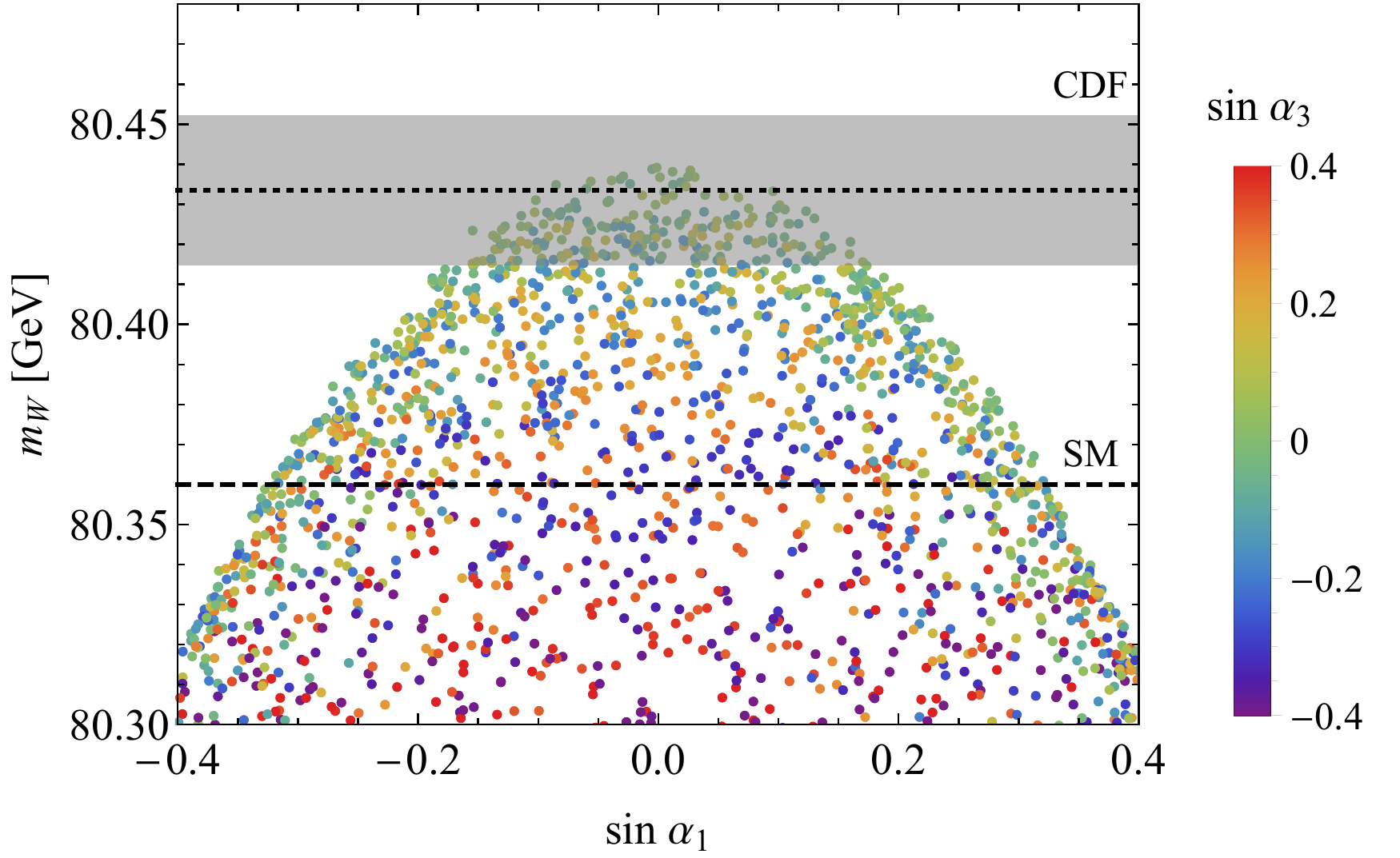}
\caption{$m_W$ and $\sin\alpha_1$ as a function of $\sin\alpha_2$ (left)  vs. $\sin\alpha_3$ (right).}
\label{mw:alphascatt}
\end{figure}

To better show the correlation between $m_W$ and the CP-even Higgs mixing angles $\alpha_i=\{\alpha_1$,$\alpha_2$,$\alpha_3\}$, we show scattering plots of $m_W$ versus $\sin\alpha_1$  in Fig.~\ref{mw:alphascatt},  by setting mixing angles $\alpha_i$ is  random values in the range $(-0.4, 0.4)$.
%
%
For other numerical inputs, we set $m_L=300$ GeV, $\eta=-200^2~{\rm GeV}^2$, $\mu=10^{-5}\,{\rm GeV}$, $v_\Delta=1\, \rm MeV$, $v_s=10^{13}\, \rm GeV$ and $m_{s}=1000\, \rm GeV$. It can be easily seen that the allowed parameter space of $m_W$ measured by  the CDF  collaboration \cite{CDF:2022hxs} are displayed as the arched areas in both panels.
However, $\sin\alpha_2$ and $\sin\alpha_3$ show the significantly different distributions. 
In the left panel, the allowed value of $\sin\alpha_2$ has a relatively uniform distribution in the range $(-0.4, 0.4)$, while the right panel shows that the allowed points for $\sin\alpha_3$ are almost smaller than $0.1$, which implies that $\sin \alpha_2$ is usually irrelevant to the corrected $m_W$. 

\subsection{The gauge-scalar interactions}
\label{int:gauge_scalar}
	
In order to express the weak eigenstates on the right side of Eq.~(\ref{mix:angle})  in terms of mass eigenstates on the left side, it's useful to get the transposed form of the orthogonal matrices, which are listed as follows,
\begin{align}
R(\beta)&=\Big[\mathcal{R}(\beta)\Big]^T=
\left(
\begin{array}{cc}
\cos \beta & \sin\beta \\
-\sin\beta   & \cos\beta
\end{array}
\right)\,,
\end{align}
\begin{eqnarray}
\begin{aligned}
V(\beta'_1,\beta'_2,\beta'_3)&=\Big[\mathcal{V}({\beta'_1,\beta'_2,\beta'_3})\Big]^T\\
&=
\left(
\begin{array}{ccc}
\cos\beta'_1\cos\beta'_2 & -\cos\beta'_1\sin\beta'_2\sin\beta'_3-\sin\beta'_1\cos\beta'_3&  -\cos\beta'_1\sin\beta'_2\cos\beta'_3+\sin\beta'_1\sin\beta'_3\\
\sin\beta'_1\cos\beta'_2   & \cos\beta'_1\cos\beta'_3-\sin\beta'_1\sin\beta'_2\sin\beta'_3& -\cos\beta'_1\sin\beta'_3-\sin\beta'_1\sin\beta'_2\cos\beta'_3\\
\sin\beta'_2   & \cos\beta'_2\sin\beta'_3  & \cos\beta'_2\cos\beta'_3
\end{array}
\right)\,,
\end{aligned}
\end{eqnarray}
\begin{eqnarray}
\begin{aligned}
U(\alpha_1,\alpha_2,\alpha_3)&=\Big[\mathcal{U}({\rm \alpha_1,\alpha_2,\alpha_3})\Big]^T\\
&=
\left(
\begin{array}{ccc}
\cos\alpha_1\cos\alpha_2 & -\cos\alpha_1\sin\alpha_2\sin\alpha_3-\sin\alpha_1\cos\alpha_3&  -\cos\alpha_1\sin\alpha_2\cos\alpha_3+\sin\alpha_1\sin\alpha_3\\
\sin\alpha_1\cos\alpha_2   & \cos\alpha_1\cos\alpha_3-\sin\alpha_1\sin\alpha_2\sin\alpha_3& -\cos\alpha_1\sin\alpha_3-\sin\alpha_1\sin\alpha_2\cos\alpha_3\\
\sin\alpha_2   & \cos\alpha_2\sin\alpha_3  & \cos\alpha_2\cos\alpha_3
\end{array}
\right)\,.
\end{aligned}
\end{eqnarray}
In the following, we list the vertices and coefficients for the corresponding interactions in Tables~\ref{fr1}, \ref{fr2}, and \ref{fr3}. For simplicity, we use the abbreviations $s_\beta(\sin\beta), c_\beta(\cos\beta)$, $\hat{s}_W(\sin\theta_w)$, and $\hat{c}_W(\cos\theta_w)$ for the mixing angle $\beta$ and the Weinberg angle $\theta_w$, respectively. In addition, the matrix elements of the matrices $U(\alpha_1,\alpha_2,\alpha_3)$ and $V(\beta'_1,\beta'_2,\beta'_3)$ are denoted as $U_{ij}, V_{ij}(i,j=1,2,3)$, respectively.

\vspace{10pt}
\begin{table}[h]
\begin{center}
{\renewcommand\arraystretch{1.5}
\begin{tabular}{|c|c||c|c|}\hline
Vertices& Coefficient &Vertices& Coefficient  \\ \hline\hline
$h W_\mu^+ W_\nu^-$  & $g^2(U_{21}\upsilon_\Delta+\frac{1}{2}U_{11}\upsilon_\phi)g_{\mu\nu}$ &$h Z_\mu Z_\nu $ &
$\frac{g^2}{2c^2_{W}}(2U_{21}\upsilon_\Delta+\frac{1}{2}U_{11}\upsilon_\phi)g_{\mu\nu}$ \\ \hline
$H W_\mu^+ W_\nu^-$  & $g^2(U_{21}\upsilon_\Delta+\frac{1}{2}U_{11}\upsilon_\phi)g_{\mu\nu}$ &$H Z_\mu Z_\nu $ &
$\frac{g^2}{2c^2_{W}}(2U_{22}\upsilon_\Delta+\frac{1}{2}U_{12}\upsilon_\phi)g_{\mu\nu}$ \\ \hline
$s W_\mu^+ W_\nu^-$  & $g^2(U_{23}\upsilon_\Delta+\frac{1}{2}U_{13}\upsilon_\phi)g_{\mu\nu}$ &$s Z_\mu Z_\nu $ &
$\frac{g^2}{2c^2_{W}}(2U_{23}\upsilon_\Delta+\frac{1}{2}U_{13}\upsilon_\phi)g_{\mu\nu}$ \\ \hline
$H^\pm W_\mu^\mp Z_\nu$  & $\left[\sqrt{2}\upsilon_\Delta c_{\beta}(-3+c^2_{2W})\frac{1}{c_{W}}+2\upsilon_\phi s_{\beta}s_{W}t_{W}\right]g_{\mu\nu} $& $G^\pm W_\mu^\mp Z_\nu$ & $\left[\sqrt{2}\upsilon_\Delta s_{\beta}(-3+c^2_{2W})\frac{1}{c_{W}}-2\upsilon_\phi c_{\beta}s_{W}t_{W}\right]g_{\mu\nu}$ \\ \hline
$H^{\pm\pm} W_\mu^\mp W_\nu^\mp$  & $\left(g^2{\upsilon_\Delta}/{\sqrt{2}}\right) g_{\mu\nu}$ & $G^\pm W_\mu^\mp A_\nu $ &
$\frac{e^2}{2s_{W}}\left(\upsilon_\phi c_\beta+\sqrt{2}\upsilon_\Delta s_\beta\right)g_{\mu\nu}$ \\ \hline
\end{tabular}}
\caption{The Higgs-gauge-gauge type interactions and the corresponding coefficients.}
\label{fr1}
\end{center}
\end{table}

\vspace{10pt}
\begin{table}[h]
\begin{center}
{\renewcommand\arraystretch{1.5}
\begin{tabular}{|c|c||c|c|}\hline
Vertices& Coefficient &Vertices& Coefficient  \\ \hline\hline
$H^{\pm\pm}H^\mp W_\mu^\mp$  & $gc_\beta(p_1-p_2)_\mu$&$H^{++}H^{--}A_\mu$&$2e(p_1-p_2)_\mu$\\ \hline
$H^{\pm\pm}G^\mp W_\mu^\mp$  & $gs_\beta(p_1-p_2)_\mu$&$H^{+}H^{-}A_\mu$&$e(p_1-p_2)_\mu$\\ \hline
$H^\pm AW_\mu^\mp$  & $-i\frac{g}{2}(V_{12} s_\beta - \sqrt{2}V_{12} c_\beta)(p_1-p_2)_\mu$&$G^{+}G^{-}A_\mu$&$e(p_1-p_2)_\mu$\\ \hline
$H^\pm HW_\mu^\mp$  & $\frac{g}{2}(-U_{12} s_\beta + \sqrt{2}U_{22}c_\beta)(p_1-p_2)_\mu$&$H^{++}H^{--}Z_\mu$ &$\frac{g}{\hat{c}_W}(\hat{c}_W^2-\hat{s}_W^2)(p_1-p_2)_\mu$\\ \hline
$H^\pm hW_\mu^\mp$  & $\frac{g}{2}(-U_{11} s_\beta+\sqrt{2}U_{21}
c_\beta)(p_1-p_2)_\mu$&$H^{+}H^{-}Z_\mu$&$\frac{g}{2\hat{c}_W}(\hat{c}_W^2-\hat{s}_W^2-c_{\beta}^2)(p_1-p_2)_\mu$\\ \hline
$H^\pm GW_\mu^\mp$  & $i\frac{g}{2}(V_{11} s_\beta - \sqrt{2} V_{21} c_{\beta})(p_1-p_2)_\mu$&$G^{+}G^{-}Z_\mu$&$\frac{g}{2\hat{c}_W}(\hat{c}_W^2-\hat{s}_W^2-s_{\beta}^2)(p_1-p_2)_\mu$\\ \hline
$H^\pm aW_\mu^\mp$  & $i\frac{g}{2}(V_{13} s_\beta - \sqrt{2} V_{23} c_{\beta})(p_1-p_2)_\mu$&$G^{+}G^{-}Z_\mu$&$\frac{g}{2\hat{c}_W}(\hat{c}_W^2-\hat{s}_W^2-s_{\beta}^2)(p_1-p_2)_\mu$\\ \hline
$H^\pm sW_\mu^\mp$  & $i\frac{g}{2}(-U_{13} s_\beta +\sqrt{2} U_{23} c_{\beta})(p_1-p_2)_\mu$&$G^{+}G^{-}Z_\mu$&$\frac{g}{2\hat{c}_W}(\hat{c}_W^2-\hat{s}_W^2-s_{\beta}^2)(p_1-p_2)_\mu$\\ \hline
$G^\pm AW_\mu^\mp$  & $-i\frac{g}{2}(V_{12} c_\beta+\sqrt{2} V_{22} s_\beta)(p_1-p_2)_\mu$&$H^{\pm}G^{\mp}Z_\mu$&$\mp\frac{g_Z}{4}s_{2\beta}(p_1-p_2)_\mu$\\ \hline
$G^\pm HW_\mu^\mp$  & $\frac{g}{2}(U_{12} c_{\beta}+\sqrt{2}U_{22} s_\beta)(p_1-p_2)_\mu$&$AHZ_\mu$&$-i\frac{g}{2\hat{c}_W}(2U_{22}V_{22}+U_{12}V_{12})(p_1-p_2)_\mu$\\ \hline
$G^\pm hW_\mu^\mp$  & $\frac{g}{2}(U_{11} c_\beta+\sqrt{2} U_{21} s_\beta)(p_1-p_2)_\mu$&$AhZ_\mu$&$-i\frac{g}{2\hat{c}_W}(2U_{21}V_{22}+U_{11}V_{12})(p_1-p_2)_\mu$\\ \hline
$G^\pm GW_\mu^\mp$  & $-i\frac{g}{2}(V_{11} c_\beta+\sqrt{2} V_{21} s_\beta)(p_1-p_2)_\mu$&$GHZ_\mu$&$-i\frac{g}{2\hat{c}_W}(2U_{22}V_{21}+U_{12}V_{11})(p_1-p_2)_\mu$\\ \hline
$G^\pm aW_\mu^\mp$  & $-i\frac{g}{2}(V_{13} c_\beta+\sqrt{2} V_{23} s_\beta)(p_1-p_2)_\mu$&$GhZ_\mu$&$-i\frac{g}{2\hat{c}_W}(2U_{21}V_{21}+U_{11}V_{11})(p_1-p_2)_\mu$\\ \hline
$G^\pm sW_\mu^\mp$  & $\frac{g}{2}(U_{13} c_\beta+\sqrt{2} U_{23} s_\beta)(p_1-p_2)_\mu$&$aHZ_\mu$&$-i\frac{g}{2\hat{c}_W}(2U_{22}V_{23}+U_{12}V_{13})(p_1-p_2)_\mu$\\ \hline
$AsZ_\mu$&$-i\frac{g}{2\hat{c}_W}(2U_{23}V_{22}+U_{13}V_{12})(p_1-p_2)_\mu$
&$ahZ_\mu$&$-i\frac{g}{2\hat{c}_W}(2U_{21}V_{23}+U_{11}V_{13})(p_1-p_2)_\mu$\\ \hline
$GsZ_\mu$&$-i\frac{g}{2\hat{c}_W}(2U_{23}V_{21}+U_{13}V_{11})(p_1-p_2)_\mu$
&$asZ_\mu$&$-i\frac{g}{2\hat{c}_W}(2U_{23}V_{23}+U_{13}V_{13})(p_1-p_2)_\mu$\\ \hline
\end{tabular}}
\caption{The Higgs-Higgs-gauge type interactions and the corresponding coefficients.}
\label{fr2}
\end{center}
\end{table}

\begin{table}[h]
\begin{center}
{\renewcommand\arraystretch{1.5}
\begin{tabular}{|c|c||c|c|}\hline
Vertices& Coefficient&Vertices& Coefficient   \\ \hline\hline
$H^{++}H^{--}W_\mu^+ W_\nu^-$  & $g^2g_{\mu\nu}$  &
$H^{++}H^{--}Z_\mu Z_\nu$  & $  \frac{g^2}{\hat{c}_W^2}(c_W^2-s_W^2)^2g_{\mu\nu}$    \\ \hline
$H^{+}H^{-}W_\mu^+ W_\nu^-$  & $\frac{g^2}{4}(5+3c_{2\beta})g_{\mu\nu}$&
$H^{+}H^{-}Z_\mu Z_\nu$   & $\frac{g^2}{8\hat{c}_W^2}\left(2+c_{4W}-4c_{2W}c_{\beta}^2+c_{2\beta}\right)g_{\mu\nu}$   \\ \hline
$G^{+}G^{-}W_\mu^+ W_\nu^-$  & $\frac{g^2}{4}(5-3c_{2\beta})g_{\mu\nu}$&
$G^{+}G^{-}Z_\mu Z_\nu$   & $\frac{g^2}{8\hat{c}_W^2}\left(2+c_{4W}-4c_{2W}s_{\beta}^2-c_{2\beta}\right)g_{\mu\nu}$\\ \hline
$HHW_\mu^+ W_\nu^-$  & $\frac{g^2}{4}(U^2_{12}+2U^2_{22})g_{\mu\nu}$&
$a a Z_\mu Z_\nu$  & $\frac{g^2}{8\hat{c}_W^2}(V^2_{13}+4V^2_{23})g_{\mu\nu}$    \\ \hline
$hhW_\mu^+ W_\nu^-$  & $\frac{g^2}{4}(U^2_{11}+2U^2_{21})g_{\mu\nu}$ &
$AAZ_\mu Z_\nu$   & $\frac{g^2}{8\hat{c}_W^2}(V^2_{12}+4V^2_{22})g_{\mu\nu}$    \\ \hline
$AAW_\mu^+ W_\nu^-$  & $\frac{g^2}{4}(V^2_{12}+2V^2_{22})g_{\mu\nu}$ &
$GGZ_\mu Z_\nu$  & $\frac{g^2}{8\hat{c}_W^2}(V^2_{11}+4V^2_{21})g_{\mu\nu}$   \\ \hline
$GGW_\mu^+ W_\nu^-$  & $\frac{g^2}{4}(V^2_{11}+2V^2_{21})g_{\mu\nu}$&
$ssZ_\mu Z_\nu$  & $\frac{g^2}{8\hat{c}_W^2}(U^2_{13}+4U^2_{23})g_{\mu\nu}$   \\ \hline
$aaW_\mu^+ W_\nu^-$  & $\frac{g^2}{4}(V^2_{13}+2V^2_{23})g_{\mu\nu}$&
$HHZ_\mu Z_\nu$  & $\frac{g^2}{8\hat{c}_W^2}(U^2_{12}+4U^2_{22})g_{\mu\nu}$   \\ \hline
$ssW_\mu^+ W_\nu^-$  &$\frac{g^2}{4}(U^2_{13}+2U^2_{23})g_{\mu\nu}$ &
$hhZ_\mu Z_\nu$  & $\frac{g^2}{8\hat{c}_W^2}(U^2_{11}+4U^2_{21})g_{\mu\nu}$   \\ \hline
$H^{++}H^{--}A_\mu Z_\nu$  & $4e\frac{g}{\hat{c}_W}(\hat{c}_W^2-\hat{s}_W^2)g_{\mu\nu}$&
$H^{++}H^{--}A_\mu A_\nu$  &  $4e^2g_{\mu\nu}$\\ \hline
$H^+H^-A_\mu Z_\nu$  & $e\frac{g}{\hat{c}_W}(\hat{c}_W^2-\hat{s}_W^2- c^2_{\beta})g_{\mu\nu}$&
$H^{+}H^{-}A_\mu A_\nu$ &$e^2g_{\mu\nu}$  \\ \hline
$G^+G^-A_\mu Z_\nu$  & $e\frac{g}{\hat{c}_W}(\hat{c}_W^2-\hat{s}_W^2-s^2_{\beta})g_{\mu\nu}$&
$G^{+}G^{-}A_\mu A_\nu$ &$e^2g_{\mu\nu}$\\ \hline
\end{tabular}}
\caption{The four-point interactions and the corresponding coefficients.}
\label{fr3}
\end{center}
\end{table}

\subsection{The ALP-scalar/neutrino interactions}
\label{int:Maj_scaneu}

We list the vertices and the corresponding coefficients for ALP-Higgs and ALP-neutrino interactions in Table~\ref{fr4}.
\begin{table}[h]
\begin{center}
{\renewcommand\arraystretch{2.4}
\begin{tabular}{|c||c|}\hline
Vertices& Coefficients \\\hline
$a^4$ & ${\frac{1}{4}{\lambda_1} V_{13}^4+\frac{1}{4}{\lambda_4}V_{13}^2 V_{23}^2+\frac{1}{4} {\lambda_5} V_{13}^2 V_{23}^2+\frac{1}{2} \lambda  V_{13}^2 V_{23} {V_{33}}+\frac{1}{4}{\lambda_2}{V_{23}^4}+\frac{1}{4}{\lambda_3}{V_{23}^4}+\frac{1}{4}}{\lambda_6} {V_{33}^4}$\\\hline
$a^3G$ &$\makecell[c]{\vspace{2pt}\lambda_1V_{11}V_{13}^3+\frac{1}{2}{\lambda_4}V_{11}V_{13}V_{23}^2+\frac{1}{2}\lambda_5 V_{11}V_{13}V_{23}^2+\lambda V_{11}V_{13} V_{23}V_{33}+\frac{1}{2} \lambda_4 V_{13}^2V_{21}V_{23}\\+\frac{1}{2}\lambda_5 V_{13}^2 V_{21}V_{23}+\frac{1}{2} \lambda V_{13}^2 V_{21} V_{33}+\frac{1}{2} \lambda  V_{13}^2 V_{23} V_{31}+\lambda_2 V_{21} V_{23}^3+\lambda_3 V_{21} V_{23}^3+\lambda_6 V_{31} V_{33}^3}$\\\hline
$a^2 h^2 $ & $\makecell[c]{\frac{1}{2} \lambda_1 U_{11}^2 V_{13}^2+\frac{1}{4} \lambda_4 U_{11}^2 V_{23}^2+\frac{1}{4} \lambda_5 U_{11}^2 V_{23}^2-\frac{1}{2} \lambda  U_{11}^2 V_{23} V_{33}+\lambda  U_{11} U_{21} V_{13} V_{33}\\-\lambda  U_{11} U_{31} V_{13} V_{23}+\frac{1}{4} \lambda_4 U_{21}^2 V_{13}^2+\frac{1}{4} \lambda_5 U_{21}^2 V_{13}^2+\frac{1}{2} \lambda_2 U_{21}^2 V_{23}^2+\frac{1}{2} \lambda_3 U_{21}^2 V_{23}^2+\frac{1}{2} \lambda U_{21} U_{31} V_{13}^2+\frac{1}{2}\lambda_6 U_{31}^2 V_{33}^2}$\\\hline
$a^2 h $ & $\makecell[c]{\lambda_1 U_{11} v_{\phi} V_{13}^2+\frac{1}{2} \lambda_4 U_{11} v_{\phi} V_{23}^2+\frac{1}{2} \lambda_5 U_{11} v_{\phi} V_{23}^2-\lambda  U_{11} v_{\phi} V_{23} V_{33}-\sqrt{2} \mu  U_{11} V_{13} V_{23}\\-\lambda  U_{11} V_{13} V_{23} v_{s}+\lambda  U_{11} V_{13} V_{33} v_\Delta+\lambda  U_{21} v_{\phi} V_{13} V_{33}+\frac{1}{\sqrt{2}}\mu  U_{21} V_{13}^2+\frac{1}{2} \lambda_4 U_{21} V_{13}^2 v_\Delta +\frac{1}{2} \lambda_5 U_{21} V_{13}^2 v_\Delta\\+\frac{1}{2} \lambda  U_{21} V_{13}^2 v_{s}+\lambda_2 U_{21} V_{23}^2 v_\Delta +\lambda_3 U_{21} V_{23}^2 v_\Delta-\lambda U_{31} v_{\phi} V_{13} V_{23}+\frac{1}{2} \lambda U_{31} V_{13}^2 v_\Delta+\lambda_6 U_{31} V_{33}^2 v_{s}}$\\\hline
$a^2 G^2$ & $\makecell[c]{\frac{3}{2} \lambda_1V_{11}^2 V_{13}^2+\frac{1}{4} \lambda_4 V_{11}^2 V_{23}^2+\frac{1}{4} \lambda_5 V_{11}^2 V_{23}^2+\frac{1}{2} \lambda  V_{11}^2 V_{23} V_{33}+\lambda_4 V_{11} V_{13} V_{21} V_{23}+\lambda_5 V_{11} V_{13} V_{21} V_{23}\\+\lambda  V_{11} V_{13} V_{21} V_{33}+\lambda  V_{11} V_{13} V_{23} V_{31}+\frac{1}{4} \lambda_4 V_{13}^2 V_{21}^2+\frac{1}{4} \lambda_5 V_{13}^2 V_{21}^2+\frac{1}{2} \lambda  V_{13}^2 V_{21} V_{31}+\frac{3}{2} \lambda_2 V_{21}^2 V_{23}^2+\frac{3}{2} \lambda_{3} V_{21}^2 V_{23}^2+\frac{3}{2} \lambda_{6} V_{31}^2 V_{33}^2}$\\\hline
$a^2 G^+G^- $ & $\makecell[c]{\lambda_1 V_{13}^2 \cos^2\beta+\frac{1}{2} \lambda_4 V_{13}^2 \sin^2\beta+\frac{1}{4}\lambda_5  V_{13}^2 \sin^2\beta+\frac{1}{\sqrt{2}}\lambda_5 V_{13} V_{23} \sin\beta \cos\beta\\+\sqrt{2} \lambda  V_{13} V_{33} \sin\beta \cos\beta+\lambda_2 V_{23}^2 \sin^2\beta+\frac{1}{2} \lambda_3 V_{23}^2 \sin^2\beta+\frac{1}{2}\lambda_{4} V_{23}^2 \cos^2\beta}$\\\hline
$a G^3$ & $\makecell[c]{\lambda_1 V_{11}^3 V_{13}+\frac{1}{2} \lambda_4 V_{11}^2 V_{21} V_{23}+\frac{1}{2} \lambda_{5} V_{11}^2 V_{21} V_{23}+\frac{1}{2} \lambda V_{11}^2 V_{21} V_{33}+\frac{1}{2} \lambda  V_{11}^2 V_{23} V_{31}\\+\frac{1}{2} \lambda_{4} V_{11} V_{13} V_{21}^2+\frac{1}{2} \lambda_5V_{11} V_{13} V_{21}^2+\lambda V_{11} V_{13} V_{21} V_{31}+\lambda_{2} V_{21}^3 V_{23}+\lambda_{3} V_{21}^3 V_{23}+\lambda_{6} V_{31}^3 V_{33}}$\\\hline
$a h^2 G$ & $\makecell[c]{\lambda_{1} U_{11}^2 V_{11} V_{13}+\frac{1}{2} \lambda_{4} U_{11}^2 V_{21} V_{23}+\frac{1}{2} \lambda_{5} U_{11}^2 V_{21} V_{23}-\frac{1}{2} \lambda  U_{11}^2 V_{21} V_{33}-\frac{1}{2} \lambda  U_{11}^2 V_{23} V_{31}\\+\lambda  U_{11} U_{21} V_{11} V_{33}+\lambda U_{11} U_{21} V_{13} V_{31}-\lambda  U_{11} U_{31}V_{11} V_{23}-\lambda  U_{11} U_{31}V_{13}V_{21}+\frac{1}{2} \lambda_{4} U_{21}^2 V_{11} V_{13}\\+\frac{1}{2} \lambda_{5} U_{21}^2 V_{11} V_{13}+\lambda_{2} U_{21}^2 V_{21} V_{23}+\lambda_{3} U_{21}^2 V_{21} V_{23}+\lambda  U_{21} U_{31} V_{11} V_{13}+\lambda_{6} U_{31}^2 V_{31} V_{33}}$\\\hline
$a\overline{\nu}\nu$ & $V_{23} m_\nu /v_\Delta$\\\hline
\end{tabular}}
\caption{The ALP-Higgs/neutrino interactions and the corresponding coefficients.}
\label{fr4}
\end{center}
\end{table}

\subsection{The gauge boson self-energies}
\label{self_energies}

The analytic expressions for gauge boson self-energies are listed as follows.
Here we use the Passarino-Veltman functions defined in Ref.~\cite{Passarino:1978jh}. 
First, the fermion-loop (F) contributions to all the two-point correlation functions are given by
\begin{align}
\Pi_{WW}^{\text{1PI}}(p^2)_{\rm F}&=\frac{g^2}{16\pi^2}N_c^f\Big[-B_4+2p^2B_3\Big](p^2,m_f,m_{f'}),\\
\Pi_{ZZ}^{\text{1PI}}(p^2)_{\rm F}&= \frac{g^2}{16\pi^2\cos^2\theta_w}N_c^f\Big[2p^2(4\sin^4\theta_wQ_f^2-4s_W^2Q_fI_f+2I_f^2)B_3-2I_f^2m_f^2B_0\Big](p^2,m_f,m_f),\\
\Pi_{\gamma\gamma}^{\text{1PI}}(p^2)_{\rm F}&=\frac{e^2}{16\pi^2}N_c^fQ_f^2\Big[8p^2B_3\Big](p^2,m_f,m_f),\\
\Pi_{Z\gamma}^{\text{1PI}}(p^2)_{\rm F}&=-\frac{eg}{16\pi^2\cos\theta_w}N_c^f\Big[2p^2(-4\sin^2\theta_wQ_f^2+2I_fQ_f)B_3\Big](p^2,m_f,m_f),
\end{align}
where~\cite{Kanemura:2012rs,Aoki:2012jj}
\begin{align}
B_3(p^2,m_1,m_2)&=-B_1(p^2,m_1,m_2)-B_{21}(p^2,m_1,m_2), \notag\\
B_4(p^2,m_1,m_2)&=-m_1^2B_1(p^2,m_2,m_1)-m_2^2B_1(p^2,m_1,m_2).
\end{align}
The scalar-loop (S1, S2), the gauge boson-loop (V), and the SM contributions to $W$-$W$ self-energy are
\begin{align}
\Pi_{WW}^{\text{1PI}}(p^2)_{\rm S1}=&-\frac{g^2}{16 \pi ^2} \bigg[B_{22}(p^2,m_{G^{+}},m_a) \left(V_{13} \cos\beta+\sqrt{2} V_{23} \sin \beta \right)^2
+B_{22}(p^2,m_{H^{+}},m_a) \left(V_{13} \sin\beta-\sqrt{2}V_{23} \cos\beta\right)^2
\nonumber\\
&+B_{22}(p^2,m_{G^{+}},m_{A}) \left(V_{12} \cos\beta +\sqrt{2}V_{22} \sin\beta \right)^2+B_{22}(p^2,m_{H^{+}},m_{A}) \left(V_{12} \sin\beta-\sqrt{2}V_{22} \cos \beta\right)^2
\nonumber\\
&+B_{22}(p^2,m_{G^{+}},m_G) \left(V_{11} \cos\beta+\sqrt{2}V_{21} \sin \beta \right)^2+B_{22}(p^2,m_{H^{+}},m_G) \left(V_{11} \sin \beta -\sqrt{2}V_{21} \cos \beta \right)^2
\nonumber\\
&+4 \sin ^2\beta B_{22}(p^2,m_{H^{++}},m_{G^{+}})+B_{22}(p^2,m_{G^{+}},m_h) \left(U_{11} \cos \beta+\sqrt{2} U_{21} \sin\beta\right)^2
\nonumber\\
&+B_{22}(p^2,m_{G^{+}},m_H) \left(U_{12} \cos \beta +\sqrt{2}U_{22} \sin \beta \right)^2+B_{22}(p^2,m_{G^{+}},m_s) \left(U_{13} \cos \beta +\sqrt{2} U_{23} \sin \beta \right)^2
\nonumber\\
&+B_{22}(p^2,m_{H^{+}},m_h) \left(\sqrt{2}U_{21} \cos\beta -U_{11} \sin \beta \right)^2+B_{22}(p^2,m_{H^{+}},m_H) \left(\sqrt{2}U_{22} \cos \beta -U_{12} \sin \beta\right)^2
\nonumber\\
&+4 \cos ^2\beta B_{22}(p^2,m_{H^{++}},m_{H^{+}})+B_{22}(p^2,m_{H^{+}},m_s) \left(\sqrt{2}U_{23} \cos \beta-U_{13} \sin\beta\right)^2\bigg];
\end{align}
\begin{align}
\Pi_{WW}^{\text{1PI}}(p^2)_{\rm S2}=&
\frac{1}{16 \pi^2}\frac{g^2}{4} \bigg[A_0(m_a) \left(V_{13}^2+2V_{23}^2\right)+A_0(m_{A}) \left(V_{12}^2+2V_{22}^2\right)+4 A_0(m_{H^{++}})
+A_0(m_G) \left(V_{11}^2+2V_{21}^2\right)
\nonumber\\
&+A_0(m_{G^{+}}) [5-3 \cos (2 \beta )]+A_0(m_h) \left(U_{11}^2+2U_{21}^2\right)+A_0(m_H) \left(U_{12}^2+2U_{22}^2\right)+A_0(m_{H^{+}})[3 \cos (2 \beta)+5]
\nonumber\\
&+A_0(m_s) \left(U_{13}^2+2U_{23}^2\right)\bigg];
\end{align}
\begin{align}
\Pi_{WW}^{\text{1PI}}(p^2)_{\rm V}=&\frac{g^4}{16 \pi ^2} \bigg[\frac{1}{16}\left(\frac{\sqrt{2}v_\Delta  \sin \beta (\cos (2\theta_w)-3)}{\cos\theta_w}-2 v_\phi \cos\beta \sin\theta_w \tan\theta_w\right)^2B_{0}(p^2,m_{G^{+}},m_{Z})
\nonumber\\
&+\frac{1}{16}\left(\frac{\sqrt{2}v_\Delta \cos\beta (\cos (2\theta_w)-3)}{\cos\theta_w}+2v_\phi\sin\beta \sin\theta_w \tan\theta_w\right)^2 B_{0}(p^2,m_{H^{+}},m_{Z})
\nonumber\\
&+\frac{1}{4} \sin^2\theta_w\left(\sqrt{2}v_\Delta \sin\beta+v_\phi\cos\beta\right)^2B_{0}(p^2,m_{G^{+}},0)+\frac{4}{2} v_\Delta^2 B_{0}(p^2,m_{H^{++}},m_{W})
\nonumber\\
&+\left(\frac{U_{12}v_\phi}{2}+U_{22}v_\Delta\right)^2B_{0}(p^2,m_{H},m_{W})+\left(\frac{U_{13}v_\phi}{2}+U_{23}v_\Delta\right)^2 B_{0}(p^2,m_{s},m_{W})
\nonumber\\
&+\left(\frac{U_{11} v_\phi}{2}+U_{21}v_\Delta\right)^2 B_{0}(p^2,m_{h},m_{W})\bigg];
\end{align}
\begin{align}
\Pi_{WW}^{\text{1PI}}(p^2)_{\rm SM}=&\frac{g^2}{16\pi^2}\Bigg\{
-\cos^2\theta_w\left[(6D-8)B_{22}+p^2(2B_{21}+2B_1+5B_0)\right](p^2,m_Z,m_W)
+(D-1)\left[\cos^2\theta_wA_0(m_Z)+A_0(m_W)\right]\notag\\
&-\sin^2\theta_w\left[(6D-8)B_{22}+p^2(2B_{21}+2B_1+5B_0)\right](p^2,0,m_W)\Bigg\}
-\frac{4g^2}{16\pi^2}(p^2-m_W^2)[\cos^2\theta_wB_0(p^2,m_Z,m_W)
\nonumber\\
&+\sin^2\theta_wB_0(p^2,0,m_W)].
\end{align}
The scalar-loop (S1, S2), the gauge boson-loop (V), and the SM contributions to $Z$-$Z$ self-energy are
\begin{align}
\Pi_{ZZ}^{\text{1PI}}(p^2)_{\rm S1}=&
-\frac{g^2}{16\pi^2\cos^2\theta_w} \bigg[(U_{11}V_{13}+2U_{21}V_{23})^2 B_{22}(p^2,m_{h},m_a)+(U_{12}V_{13}+2U_{22}V_{23})^2 B_{22}(p^2,m_{H},m_a)
\nonumber\\
&+(U_{13}V_{13}+2U_{23}V_{23})^2B_{22}(p^2,m_a,m_s)+(U_{11}V_{12}+2U_{21}V_{22})^2 B_{22}(p^2,m_h,m_{A})
\nonumber\\
&+(U_{12}V_{12}+2U_{22}V_{22})^2 B_{22}(p^2,m_{H},m_{A})+(U_{13}V_{12}+2U_{23}V_{22})^2B_{22}(p^2,m_s,m_{A})
\nonumber\\
&+4 \left(\cos ^2\theta_w-\sin ^2\theta_w\right)^2B_{22}(p^2,m_{H^{++}},m_{H^{++}})+(U_{11}V_{11}+2U_{21}V_{21})^2B_{22}(p^2,m_{h},m_G)
\nonumber\\
&+(U_{12}V_{11}+2U_{22}V_{21})^2B_{22}(p^2,m_{H},m_G)+(U_{13}V_{11}+2U_{23}V_{21})^2B_{22}(p^2,m_s,m_G)
\nonumber\\
&+2 \sin ^2\beta\cos ^2\beta B_{22}(p^2,m_{H^{+}},m_{G^{+}})+\left(-\sin ^2\beta-\sin^2\theta_w+\cos^2\theta_w\right)^2B_{22}(p^2,m_{G^{+}},m_{G^{+}})
\nonumber\\
&+\left(-\cos^2\beta-\sin^2\theta_w+\cos ^2\theta_w\right)^2B_{22}(p^2,m_{H^{+}},m_{H^{+}}) \bigg];
\end{align}
\begin{align}
\Pi_{ZZ}^{\text{1PI}}(p^2)_{\rm S2}=&\frac{1}{16 \pi^2}\frac{g^2}{ \cos^2\theta_w} \bigg[\frac{1}{4} A_0(m_a)\left(V_{13}^2+4V_{23}^2\right)+\frac{1}{4} A_0(m_{A}) \left(V_{12}^2+4V_{22}^2\right)+2A_0(m_{H^{++}}) \left(\cos ^2\theta_w-\sin ^2\theta_w\right)^2
\nonumber\\
&+\frac{1}{4}A_0(m_G) \left(V_{11}^2+4V_{21}^2\right)+\frac{1}{4}A_0(m_{G^{+}}) \left[-4\sin^2\beta \cos (2\theta_w)-\cos (2\beta)+
\cos (4\theta_w)+2\right]
\nonumber\\
&+\frac{1}{4}A_0(m_h) \left(U_{11}^2+4U_{21}^2\right)+\frac{1}{4}A_0(m_{H})\left(U_{12}^2+4U_{22}^2\right)+\frac{1}{4}A_0(m_{H^{+}}) \left[-4 \cos^2\beta \cos(2\theta_w)+\cos (2\beta )+\cos (4\theta_w)+2\right]
\nonumber\\
&+\frac{1}{4}A_0(m_s) \left(U_{13}^2+4U_{23}^2\right)\bigg];
\end{align}
\begin{align}
\Pi_{ZZ}^{\text{1PI}}(p^2)_{\rm V}=&\frac{g^4}{16 \pi ^2} \Bigg[\frac{2}{16}B_0(p^2,m_{G^{+}},m_W) \left(\frac{\sqrt{2}v_\Delta \sin\beta (\cos (2 \theta_w)-3)}{\cos\theta_w}-2v_\phi \cos\beta \sin\theta_w \tan\theta_w\right)^2
\nonumber\\
&+\frac{2}{16}B_0(p^2,m_{H^{+}},m_W) \left(\frac{\sqrt{2}v_\Delta \cos\beta (\cos (2\theta_w-3)}{\cos\theta_w}+2v_\phi \sin\beta
\sin\theta_w \tan\theta_w\right)^2+\frac{1}{\cos^4\theta_w}\times
\nonumber\\
&\bigg[\left(\frac{U_{11}v_\phi}{2}+2U_{21}v_\Delta\right)^2B_0(p^2,m_{h},m_{Z})+\left(\frac{U_{12}v_\phi}{2}+2U_{22}v_\Delta \right)^2 B_0(p^2,m_{H},m_{Z})
\nonumber\\
&+\left(\frac{U_{13}v_\phi}{2}+2U_{23}v_\Delta\right)^2B_0(p^2,m_s,m_{Z})\bigg]\Bigg];
\end{align}
\begin{align}
\Pi_{ZZ}^{\text{1PI}}(p^2)_{\rm SM}=&\frac{g^2}{16\pi^2\cos^2\theta_w}\Bigg\{
-\cos^4\theta_w\left[(6D-8)B_{22}+p^2(2B_{21}+2B_1+5B_0)\right](p^2,m_W,m_W)
+2(D-1)\cos^4\theta_wA_0(m_W)\Bigg\}
\nonumber\\
&-\frac{4g^2}{16\pi^2}\cos^2\theta_w(p^2-m_Z^2)B_0(p^2,m_W,m_W).
\end{align}
The scalar-loop (S1, S2), the gauge boson-loop (V), and the SM contributions to $\gamma$-$\gamma$ self-energy are
\begin{align}
\Pi_{\gamma\gamma}^{\text{1PI}}(p^2)_{\rm S1}=-\frac{e^2}{16 \pi^2}\bigg[16B_{22}(p^2,m_{H^{++}},m_{H^{++}})+4B_{22}(p^2,m_{G^{+}},m_{G^{+}})+4
B_{22}(p^2,m_{H^{+}},m_{H^{+}})\bigg];
\end{align}
\begin{align}
\Pi_{\gamma\gamma}^{\text{1PI}}(p^2)_{\rm S2}=\frac{e^2}{16 \pi ^2}\bigg[8A_0(m_{H^{++}})+2A_0(m_{G^{+}})+2A_0(m_{H^{+}})\bigg];
\end{align}
\begin{align}
\Pi_{\gamma\gamma}^{\text{1PI}}(p^2)_{\rm V}=\frac{e^4}{\left(16 \pi ^2\right)} \frac{1}{2\sin ^2\theta_w}B_{0}(p^2,m_{G^{+}},m_{W}) \left(\sqrt{2}v_\Delta\sin \beta+v_\phi\cos\beta\right)^2;
\end{align}
\begin{align}
\Pi_{\gamma\gamma}^{\text{1PI}}(p^2)_{\rm SM}=&-\frac{e^2}{16\pi^2}\Big[(6D-8)B_{22}(p^2,m_W,m_W)+p^2(2B_{21}+2B_1+5B_0)(p^2,m_W,m_W)\notag\\
&-2(D-1)A_0(m_W)-2m_W^2B_0(p^2,m_{G^+},m_W)\Big]-\frac{4e^2}{16\pi^2}p^2B_0(p^2,m_W,m_W).
\end{align}
The scalar-loop (S1, S2), the gauge boson-loop (V), and the SM contributions to $Z$-$\gamma$ self-energy are
\begin{align}
\Pi_{Z\gamma}^{\text{1PI}}(p^2)_{\rm S1}=&\frac{1}{16 \pi^2}\frac{eg}{\cos\theta_w} \bigg[8 \left(\cos ^2\theta_w-\sin^2\theta_w\right)
B_{22}(p^2,m_{H^{++}},m_{H^{++}})+2\left(-\sin^2\beta-\sin^2\theta_w+\cos^2\theta_w\right)B_{22}(p^2,m_{G^{+}},m_{G^{+}})
\nonumber\\
&+2\left(-\cos^2\beta-\sin^2\theta_w+\cos^2\theta_w\right)B_{22}(p^2,m_{H^{+}},m_{H^{+}}) \bigg];
\end{align}
\begin{align}
\Pi_{Z\gamma}^{\text{1PI}}(p^2)_{\rm S2}=&-\frac{1}{16\pi^2}\frac{eg}{\cos\theta_w} \bigg[4A_0(m_{H^{++}}) \left(\cos ^2\theta_w-\sin^2\theta_w\right)+A_0(m_{G^{+}}) \left(-\sin^2\beta-\sin^2\theta_w+\cos^2\theta_w\right)
\nonumber\\
&+A_0(m_{H^{+}}) \left(-\cos^2\beta-\sin ^2\theta_w+\cos^2\theta_w\right)\bigg];
\end{align}
\begin{align}
\Pi_{Z\gamma}^{\text{1PI}}(p^2)_{\rm V}=-\frac{1}{16\pi^2}\frac{e^2 g^2}{4 \sin\theta_w}B_{0}(p^2,m_{G^{+}},m_{W}) \left(\sqrt{2}v_\Delta\sin\beta+v_\phi\cos \beta\right) \left(\frac{\sqrt{2}v_\Delta \sin\beta(\cos (2\theta_w)-3)}{\cos\theta_w}-2v_\phi\cos\beta \sin \theta_w \tan\theta_w\right);
\end{align}
\begin{align}
\Pi_{Z\gamma}^{\text{1PI}}(p^2)_{\rm SM}=&\frac{eg}{16\pi^2\cos\theta_w}\Big[\cos^2\theta_w(6D-8)B_{22}(p^2,m_W,m_W)+\cos^2\theta_wp^2(2B_{21}
+2B_1+5B_0)(p^2,m_W,m_W)\notag\\
&-2\cos^2\theta_w(D-1)A_0(m_W)+2m_W^2\left(\sin^2\theta_w+\sin^2\beta\right)B_0(p^2,m_{G^+},m_W)\Big]
\notag\\
&+\frac{4eg}{16\pi^2\cos\theta_w}\cos^2\theta_w\left(p^2-\frac{1}{2}m_Z^2 \right)B_0(p^2,m_W,m_W).
\end{align}


\end{document}